\documentclass[a4paper,fleqn,usenatbib]{mnras}
\usepackage{newtxtext,newtxmath}
\usepackage{appendix,apptools}
\usepackage{float}
\usepackage{graphicx}	
\usepackage{amsmath}	
\usepackage{amssymb}	
\usepackage{amsfonts}      
\usepackage{gensymb}        
\usepackage{upgreek}
\usepackage{adjustbox}
\usepackage{subfigure}
\usepackage[bottom]{footmisc}
\usepackage{multirow}
\usepackage[flushleft]{threeparttable}
\usepackage{verbatim}
\begin{document}
\title[SUPERB II]{The SUrvey for Pulsars and Extragalactic Radio
       Bursts II: New FRB discoveries and their follow-up}
       
\author[S. Bhandari et al.]{\vspace{-10ex} S. Bhandari,$^{1,2,3}$\thanks{Email: shivanibhandari58@gmail.com}
E. F. Keane,$^{4,1,2}$ 
E. D. Barr,$^{9,1,2}$ 
A. Jameson,$^{1,2}$
E. Petroff,$^{5,1,2,3}$ 
\newauthor
S. Johnston,$^3$
M. Bailes,$^{1,2}$ 
N. D. R. Bhat,$^{2,10}$ 
M. Burgay,$^{11}$ 
S. Burke-Spolaor,$^{6,7}$ 
 \newauthor
M. Caleb,$^{2,12}$  
R. P. Eatough,$^{9}$ 
C. Flynn,$^{1,2}$
J. A. Green,$^{3}$ 
F. Jankowski,$^{1,2}$ 
M. Kramer,$^{9,15}$ 
\newauthor
V.Venkatraman Krishnan,$^{1,2}$ 
V. Morello,$^{9,1}$
A. Possenti,$^{11}$ 
B. Stappers,$^{15}$
C. Tiburzi,$^{16}$ 
\newauthor
W. van Straten,$^{1,17}$ 
I. Andreoni,$^{1,2,8}$ 
T. Butterley,$^{29}$
P. Chandra,$^{13}$ 
J. Cooke,$^1$ 
A.Corongiu,$^{11}$
\newauthor
D.M. Coward,$^{21}$
V. S. Dhillon,$^{15,27}$
R. Dodson,$^{24}$
L. K. Hardy,$^{14}$ 
E.J. Howell,$^{21}$
\newauthor
P. Jaroenjittichai,$^{30}$
A. Klotz,$^{22,23}$
S. P. Littlefair,$^{15}$
T. R. Marsh,$^{28}$
M. Mickaliger,$^{15}$
\newauthor
T. Muxlow,$^{15}$ 
D. Perrodin,$^{11}$
T. Pritchard,$^{1}$ 
U. Sawangwit,$^{30}$
T. Terai,$^{25}$
N. Tominaga,$^{19,31}$
\newauthor
P. Torne,$^{9}$
T. Totani,$^{20}$
A. Trois,$^{11}$
D.Turpin,$^{22,23}$
Y. Niino,$^{26}$
R. W. Wilson,$^{29}$
\newauthor
The ANTARES Collaboration.\thanks{Full author list and affiliations included at the end of the paper}
}
\date{Accepted XXX. Received YYY; in original form ZZZ}
\pubyear{2017}

\label{firstpage}
\pagerange{\pageref{firstpage}--\pageref{lastpage}}
\maketitle
\begin{abstract}
We report the discovery of four Fast Radio Bursts (FRBs) in the ongoing SUrvey for Pulsars and Extragalactic Radio Bursts (SUPERB) at the Parkes Radio Telescope: FRBs~150610, 151206, 151230 and 160102. Our real-time discoveries have enabled us to conduct extensive, rapid multi-messenger follow-up at 12 major facilities sensitive to radio, optical, X-ray, gamma-ray photons and neutrinos on time scales ranging from an hour to a few months post-burst. No counterparts to the FRBs were found and we provide upper limits on afterglow luminosities. None of the FRBs were seen to repeat. Formal fits to all FRBs show hints of scattering while their intrinsic widths are unresolved in time. FRB~151206 is at low Galactic latitude, FRB~151230 shows a sharp spectral cutoff, and FRB~160102 has the highest dispersion measure (DM = $2596.1\pm0.3$~pc~cm$^{-3}$) detected to date. Three of the FRBs have high dispersion measures (DM >$1500$~pc~cm$^{-3}$), favouring a scenario where the DM is dominated by contributions from the Intergalactic Medium. 
The slope of the Parkes FRB source counts distribution with fluences $>2$~Jy~ms is $\alpha=-2.2^{+0.6}_{-1.2}$ and still consistent with a Euclidean distribution ($\alpha=-3/2$). We also find that the all-sky rate 
is $1.7^{+1.5}_{-0.9}\times10^3$FRBs/($4\pi$~sr)/day above $\sim2~\rm{Jy}~\rm{ms}$ and there is currently no strong evidence for a latitude-dependent FRB sky-rate. 
\end{abstract}

\section{Introduction}
High-time resolution studies of the radio Universe have led to the
discovery of Fast Radio Bursts (FRBs). First seen in 2007 in archival
Parkes radio telescope data \citep{Lorimer}, FRBs have dispersion
measures (DMs) which can exceed the Milky Way contribution by more
than an order of magnitude \citep{frbcat} and typically have durations
of a few milliseconds. In the past couple of years the discovery rate
has accelerated --- including those reported here, there are now 31
FRBs known --- which include discoveries from the Green Bank
Telescope (GBT), the Parkes radio telescope, the Arecibo Observatory, the upgraded Molonglo
synthesis telescope (UTMOST) and the Australian SKA Pathfinder (ASKAP)
\citep{Lorimer,Keane,Thornton,Burke,Spitler,2014b,Vikram,Champion,GBT,keane2016host,RaviScience,FRB150215,UTMOSTFRBs,ASKAP}.

The origin of these bursts is currently unknown, with leading theories
suggesting giant flares from magnetars \citep{Thornton,Pen}, compact
objects located in young expanding supernovae \citep{Connor,antony}
and supergiant pulses from extragalactic neutron stars
\citep{cordesNS} as possible progenitors. Other theories involve
cataclysmic models including neutron star mergers \citep{Totani} and
``blitzars'' occurring when a neutron star collapses to a black hole
\citep{Falcke}.

Independent of the physical mechanism/process, an FRB may leave an
afterglow through interaction with the surrounding
medium. \citet{Yi2014} have estimated FRB afterglow luminosities,
using standard GRB afterglow models in radio, optical and X-ray bands,
assuming a plausible range of total kinetic energies and redshifts.
\citet{Maxim} have discussed possible electromagnetic counterparts
for FRBs; searching for such counterparts is thus one strategy for
localising FRB host galaxies. 
\citet{VLAlocalisation} directly
localised the repeating FRB 121102 \citep{spitler2016repeating} using the Karl
G. Jansky Very Large Array Telescope (VLA) and identified its host to be a
dwarf galaxy at a redshift $z\sim0.2$ \citep{Host}. The host is
co-located with a persistent variable radio source. Additionally, the
radio follow-ups of FRB 131104 \citep{radiocurve} and FRB 150418
\citep{keane2016host,Simon} have shown the presence of variable radio emission from
Active Galactic Nuclei (AGN) in the fields of FRBs.

The SUrvey for Pulsars and Extragalactic Radio Bursts (SUPERB) is
currently ongoing at the Parkes radio telescope and is described in
detail in \citet{SUPERB1}, hereafter Paper 1.  Initial results from
the SUPERB survey have already been published elsewhere --- this
includes investigations into radio frequency interference (RFI) at the
Parkes site~\citep{SUPERB}, the discovery of
FRB~150418~\citep{keane2016host} and the discovery of new pulsars
(Paper 1). Here we report further results from the survey, in
particular the discovery of four new FRBs ---150610, 151206, 151230
and 160102 --- as well as the multi-messenger
follow-up of the four FRBs.In \S\ref{observation}, we provide an overview of the
observations and techniques for the FRB search. Next we present the
new FRB discoveries and their properties in \S\ref{discovery}. FRB
multi-messenger follow-up observations and their results are described
in \S\ref{follow-up}. Finally, in \S\ref{end} and \S\ref{conclusion}
we present our conclusions and discuss the implications of our
results.
\section{Observations and Techniques} \label{observation}

The full details of the SUPERB observing system and analysis setup can
be found in Paper 1; here we briefly summarise the key points relevant
to this work. Real-time searches are conducted for both transient and
periodic signals in the incoming data. These data are also searched
offline through a more rigorous process which operates slower than
real time. These two streams are called the ``Fast" (F) and
``Thorough" (T) pipelines, respectively. For the single pulse
pipeline, data are acquired in the form of a time, frequency and
total intensity matrix. These are fed to the transient detection
pipeline,
\textsc{heimdall}\footnote{\url{https://sourceforge.net/projects/heimdall-astro/}},
which applies sliding boxcar filters of various widths and performs a
threshold search. This produces candidate detections that are
classified as FRBs if they meet the following criteria:
\begin{equation}
  \centering
  \label{eq:criteria}
  \begin{split}
    \rm DM \geq 1.5 \times \rm DM_{\rm Galaxy}\, \\
    \rm S/N \geq 8\, \\
    \rm N_{\rm beams,adj}  \leq 4\, \\
    \rm W  \leq 262.14\rm~ms\, \\
    \rm N_{\rm events}(t_{\rm obs} - 2\rm s\rightarrow t_{\rm obs} + 2\rm s)\; \leq 5\, \\
  \end{split}
\end{equation}
where DM and DM$_{\rm Galaxy}$ are the dispersion measures of the
candidate and the Milky Way contribution along the line of sight,
respectively. The latter is estimated using the NE2001 model
\citep{Cordes}. S/N is the peak signal-to-noise ratio of the candidate,
N$_{\rm beams,adj}$ is the number of adjacent beams in which the
candidate is detected and W is the width of the boxcar. The final
criterion measures the number of candidates detected within a 4-second
window centred on the time of occurrence of the pulse. If there are
too many candidates in a time region around the candidate of interest,
it is flagged as RFI. These criteria are followed by the T-pipeline, and
for the purposes of keeping the processing to real-time, for the F-pipeline, 
we raise the detection threshold to $\rm S/N \geq 10$ 
and only search for pulses with widths $\rm W  \leq 8.192\;\rm ms$.
When a candidate meets these criteria, an alert
email is issued and an astronomer evaluates a series of diagnostic
plots to determine the validity of the candidate. If the candidate is
deemed credible, multi-wavelength follow-up is triggered.  Upon
detection of a candidate matching the above criteria, 8-bit
full-Stokes data are saved to disk for further offline processing.

\section{FRB discoveries} \label{discovery}
\begin{figure*}
   \centering
    \includegraphics[width=\columnwidth, trim = 0mm 10mm 0mm 20mm, clip]{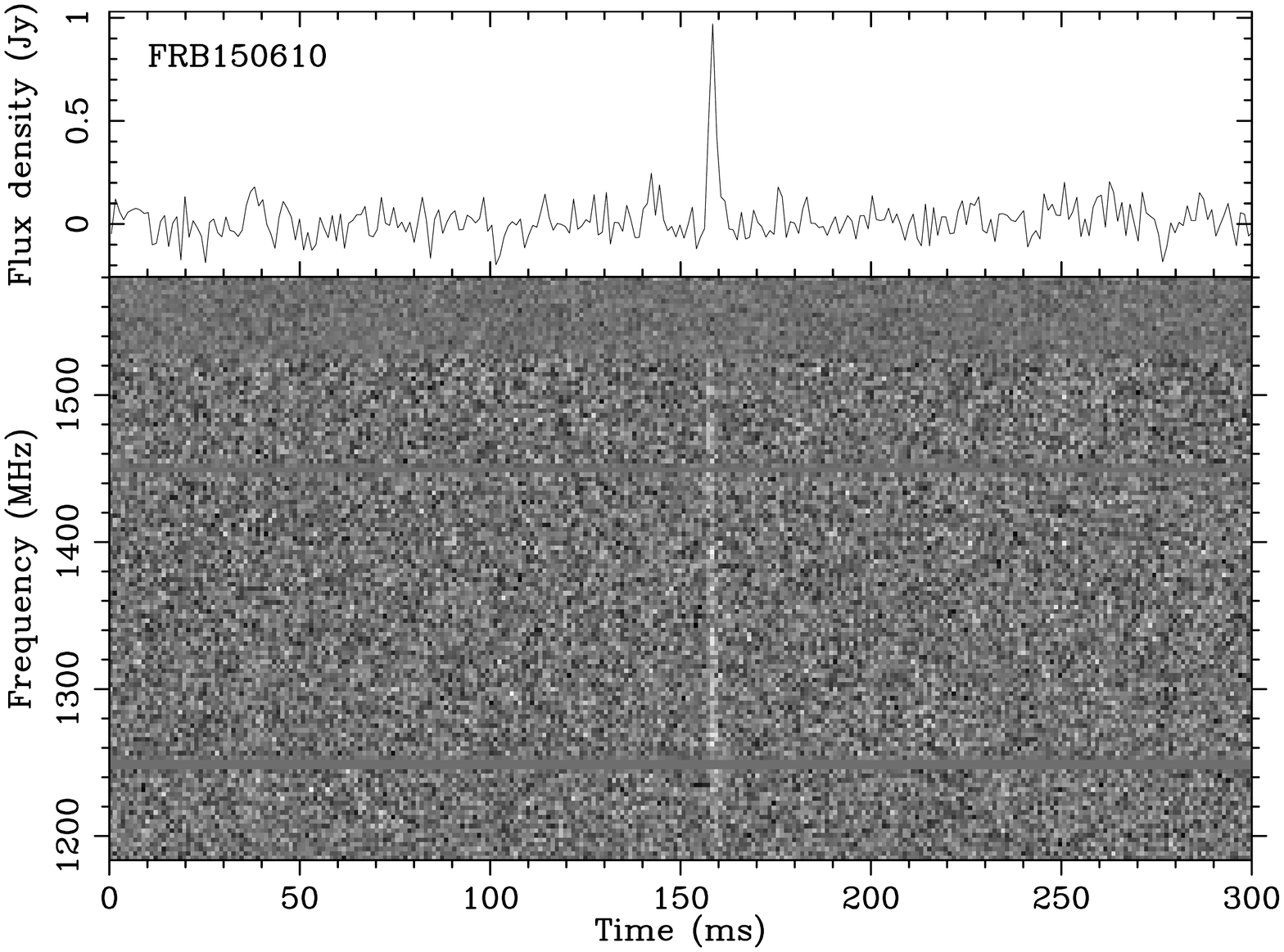}
   \includegraphics[width=\columnwidth,trim = 0mm 10mm 0mm 20mm, clip]{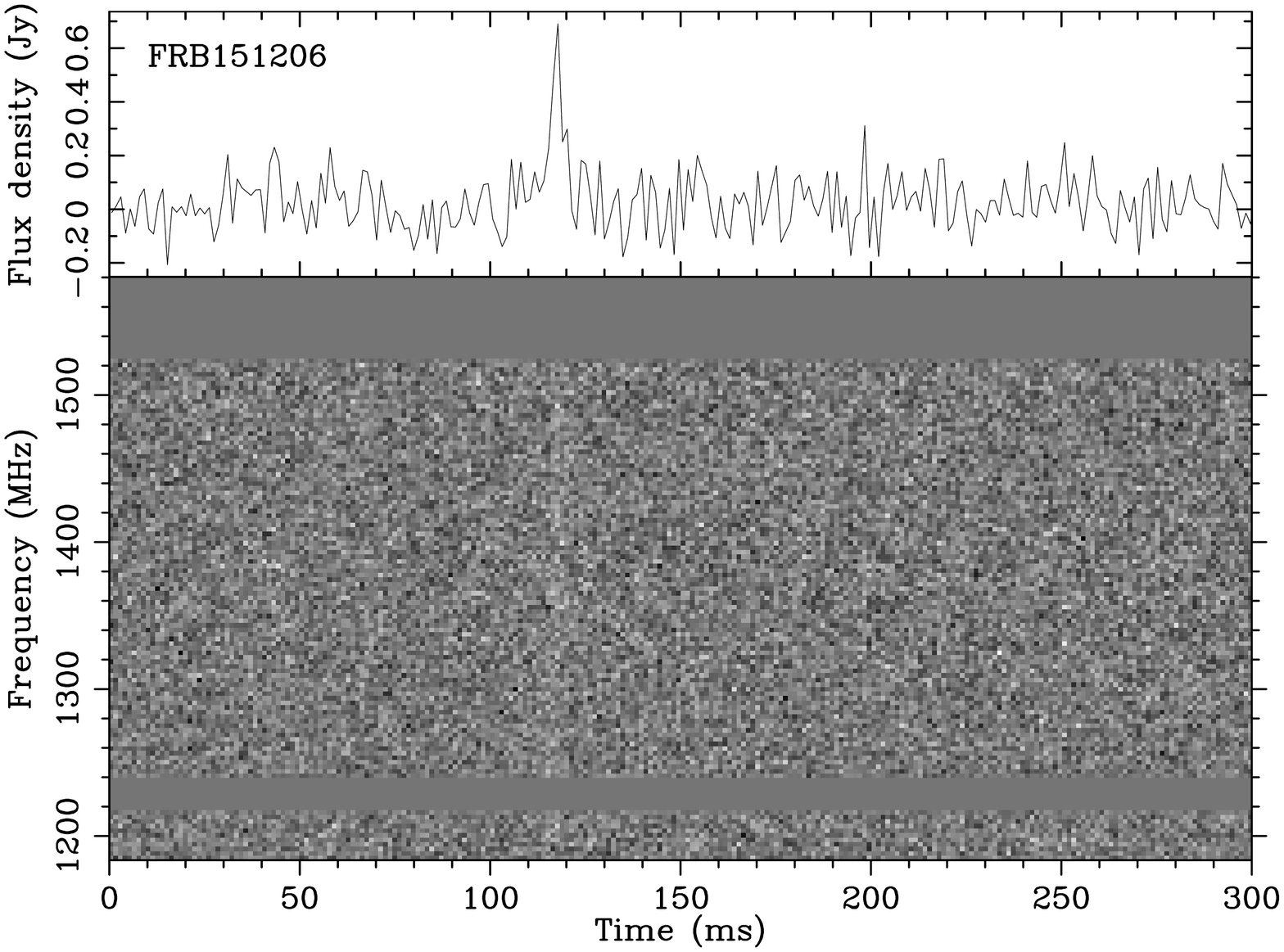}
   \includegraphics[width= \columnwidth,trim = 0mm 10mm 0mm 20mm, clip]{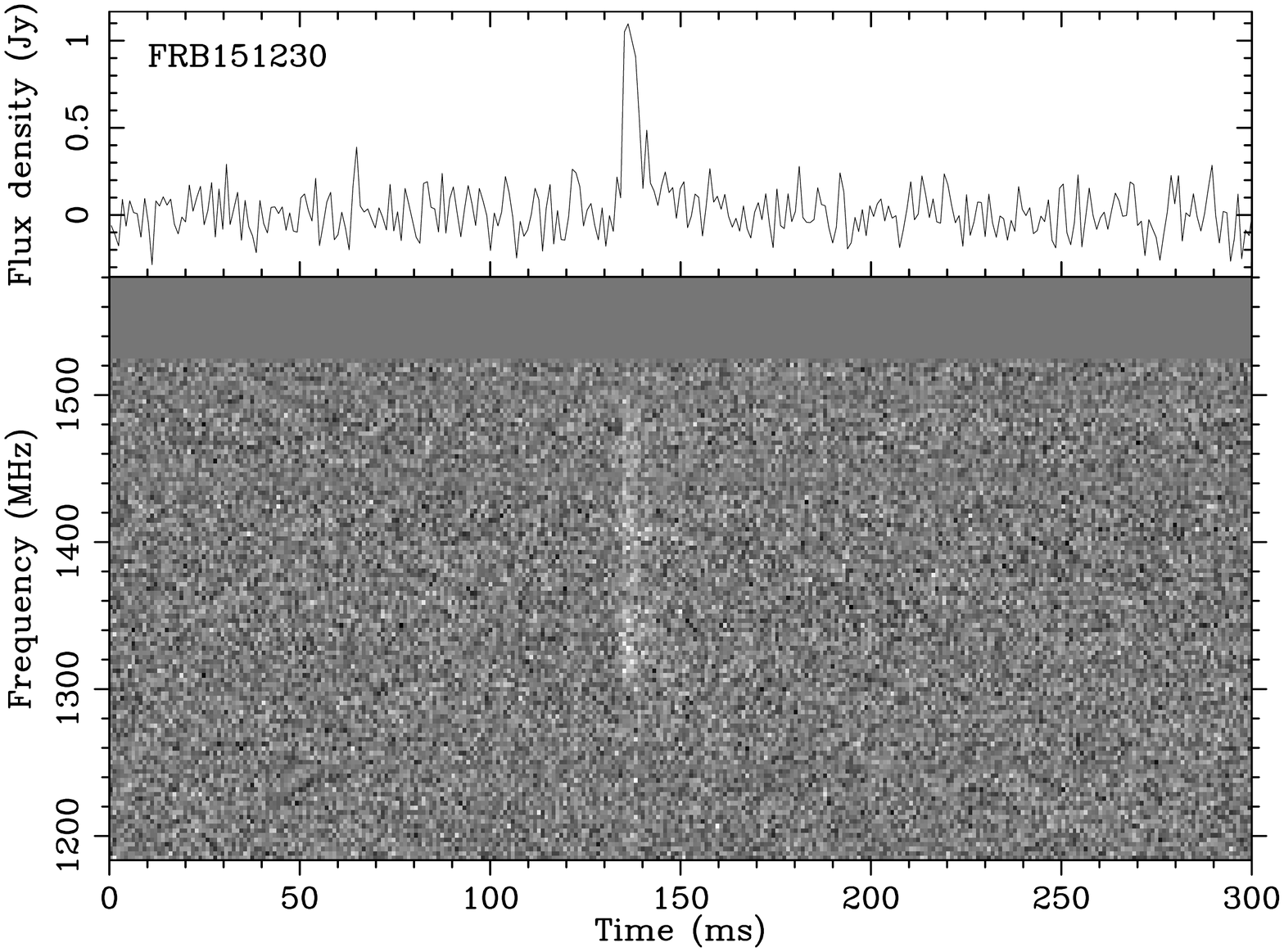}
   \includegraphics[width=\columnwidth,trim = 0mm 10mm 0mm 20mm, clip]{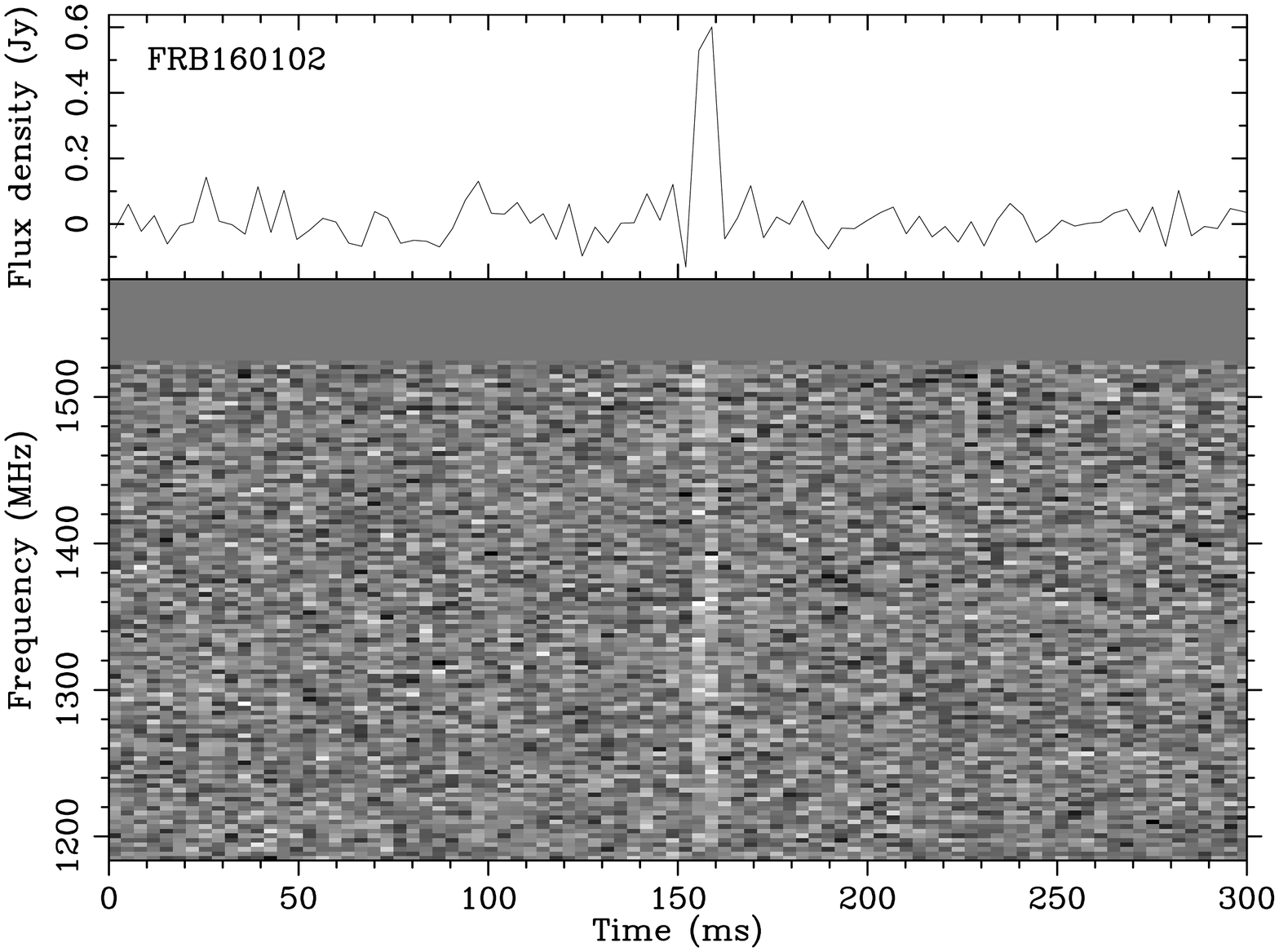}
\caption{The pulse profiles of the four new FRBs de-dispersed to
     their best-fitting DM values: clock-wise from top left FRB
     150610, FRB 151206, FRB 160102 and FRB 151230. The top panel shows the time series, frequency averaged to one channel and the bottom \
     panel shows the spectrum of the pulse. The data have been time averaged to 1 ms, 0.6 ms, 0.8 ms and  0.5 ms per sample for FRB 150610, FRB 151206, FRB 160102 and FRB 151230 respectively. The flux density scale in the upper panel of individual pulses is derived from the radiometer equation. See table \ref{spects} for
     the dispersion smearing times within a single channel for each FRB.}
   \label{figure:All_FRBs}
   \end{figure*}

	The individual pulse profiles for the FRBs are shown in
	Fig~\ref{figure:All_FRBs} and Table~\ref{spects} presents their
	measured and derived properties. The FRBs were detected in single 
	beam of the Parkes multi-beam receiver. Each FRB has a positional uncertainty 
	with a radius of 7.5$\arcmin$. The inferred properties including redshift,
        energy, co-moving and luminosity distance are derived using
        the YMW16 model \citep{YMW16} of the electron density in the
        Milky Way. Our results are consistent within the 
        uncertainties if we adopt the NE2001 model \citep{Cordes}
        instead. To measure the scattering properties of the bursts,
        the procedure adopted in \citet{Champion} was applied. 
        The resulting scattering time was scaled to a standard frequency of 1 GHz, 
        using a spectral index of $-4$. In the fitting process, we varied 
        the assumed intrinsic width of the burst and find in all cases that the best fit is 
        given by a burst duration that is solely determined by a 
        combination of DM-smearing across the filterbank channels and 
        interstellar scattering. Hence, due to the high DM of the FRBs reported here, all 			four FRBs are unresolved in width. 
        We note that the estimated isotropic energies of the FRBs at source
       had an incorrect redshift correction in \citet{simu}. The FRBs
       analysed at that time were mainly at redshifts $z < 1$, and the
       conclusions of the paper are unaffected. In this paper, three of the
       reported FRBs have DM $> 1500$ pc cm$^{-3}$, for which cosmological
       effects become important.
       We follow \citet{hogg9905116distance} and
       estimate the in-band intrinsic energies of the FRBs as:
	\begin{equation}
	E(J)= \frac{\mathcal{F}_{\rm obs} \times BW \times 4 \pi D_{\rm L}^{2}
	 \times 10^{-29}}{(1+z)^{1+\alpha}}
	\end{equation}
	where $\mathcal{F}_{\rm obs}$ is the observed fluence for FRBs in Jy ms, $BW$ is the bandwidth at Parkes in Hz, 
	$D_{\rm L}$ is the luminosity distance in meters, $z$ is the inferred redshift of FRBs and $\alpha$ is the spectral index of the
	source. Note that the denominator incorporates both the $k$-correction for the spectral index and the time-dilation correction.
	Since we generally assume the spectral index to be flat and thus $\alpha=0$, there is no $k$-correction in practice.
   
    \begin{table*}
   \centering
    \caption{\small{The observed and inferred (model-dependent)
        properties for FRBs 150610, 151206, 151230 and 160102. The model-dependent properties are derived using
        the YMW16 model \citep{YMW16} of the electron density in the
        Milky Way. For the cosmological parameters we use CosmoCalc \citep{Cosmocalc}, adopting
        $H_{0}$ = $69.9~\rm$ km $\rm s^{-1}$ Mpc$^{-1}$,
        $\Omega_{\mathrm{M}} = 0.286$ and $\Omega_{\upLambda} = 0.714$. The error in the isotropic energy estimate 
        is dominated by the error in the fluence.}}
      \label{spects}
      \begin{adjustbox}{max width=\textwidth} 
      \begin{tabular}{|c|c|c|c|c|}
      \hline
      \textbf{FRB YYMMDD} & \textbf{FRB 150610} & \textbf{FRB 151206}  & \textbf{FRB 151230} & \textbf{FRB 160102} \\ 
      \hline
      &  & \textbf{Measured Properties} & & \\
      \hline
      Event time at 1.4~GHz UTC & 2015-06-10 05:26:59.396 & 2015-12-06 06:17:52.778  &2015-12-30 16:15:46.525 & 2016-01-02 08:28:39.374 \\
      Parkes beam number & 02 &03 & 04 &13  \\
      RA, DEC (J2000) &10:44:26, $-$40:05:23 &19:21:25, $-$04:07:54 & 09:40:50, $-$03:27:05 &  22:38:49,  $-$30:10:50  \\
      ($\ell$, $b$) & $278.0\degree$, 16.5$\degree$  & 32.6$\degree$, $-$8.5$\degree$  & 239.0$\degree$, 34.8$\degree$ & 18.9$\degree$, $-$60.8$\degree$\\
      Signal to noise ratio, (S/N) & 18 & 10 & 17 & 16 \\
      Dispersion measure, DM (pc cm$^{-3}$) & 1593.9$\pm$0.6 & 1909.8$\pm$0.6 & 960.4$\pm$0.5  & 2596.1$\pm$0.3  \\
      Scattering time at 1~GHz (ms) & 3.0$\pm$0.9 & 11$\pm$2 &18$\pm$6 &4$\pm$1  \\
      Measured width, W50 (ms) & 2.0$\pm$1.0 & 3.0$\pm$0.6 & 4.4$\pm$0.5  & 3.4$\pm$0.8  \\
      Instrumental dispersion smearing (ms) &  2.0 &2.3 & 1.2 & 3.2 \\
      Observed peak flux density, S$_{\rm peak}$ (Jy)& $0.7\pm$0.2 & 0.30$\pm$0.04 & 0.42$\pm$0.03 &  0.5$\pm$0.1 \\
      Measured fluence (Jy ms) &  $>$1.3$\pm$0.7 & $>$0.9$\pm$0.2 & $>$1.9$\pm$0.3 & $>$1.8$\pm$0.5 \\
      \hline
      & & \textbf{Model-dependent properties}  & & \\
      \hline
      DM$_{\rm Gal}$ (pc cm$^{-3}$)& $\sim$122 & $\sim$160 & $\sim$38 & $\sim$13 \\
      Max. inferred $z$             &      1.2 & 1.5  &  0.8  & 2.1\\
      Max. comoving distance (Gpc)       &        3.9 &  4.3   &   2.7  &   5.5 \\
      Max. luminosity distance (Gpc)     &         8.6  & 10.6  &  4.8  &  17.2 \\
      Max. isotropic energy (10$^{33}$ J) &        1.8$\pm$1.0  &  1.7$\pm$0.4   &   1.0$\pm$0.2  &  7.0$\pm$2.0\\
      Average luminosity (10$^{36}$ W) & 0.9$\pm$0.7 & 0.6$\pm$0.2 &  0.2$\pm$0.04 & 2.0$\pm$0.7\\
      \hline
    \end{tabular}
  \end{adjustbox}
  \end{table*}
\textbf{\textit{FRB 150610}} was not detected in the F-pipeline. The
reason for this was the final selection criterion described in
Equation~\ref{eq:criteria}. At the time of observation, the number of
events detected in a 4-second window did not make a distinction
\textit{by beam} and as such was overly harsh. In this case, one beam
(beam 10) had a large number of RFI events in the time window, which
resulted in all other (unrelated) beams being flagged. This criterion
has since been corrected in the F-pipeline. FRB 150610 was discovered
in the T-pipeline which makes less severe cuts to generated
candidates. Since this burst was found in the offline processing, no
prompt follow-up observations could be performed upon detection.
The burst is slightly scattered but unresolved.\footnote{In the lowest subbands a second peak is visible, 
but statistical tests suggest that it is not 
significant and caused by noise fluctuations.}
We determine the frequency dependence of the observed 
dispersion, t$_{\rm delay}$ $\propto$ DM $\times$ $\nu^{-\beta}$, to be 
$\beta=2.000 \pm 0.008$, perfectly consistent with a cold-plasma law. 

\textbf{\textit{FRB 151206}} fell \textit{just} between search trials
in the F pipeline, placing it slightly below the detection
threshold. However the T-pipeline (which samples DM parameter space
more completely) identified it soon after. As a result the full-Stokes
data were not retained and no polarisation information is available. The burst is 
unresolved and slightly scattered. The limited signal-to-noise ratio prevents a fit for the 
DM index. The trigger was issued only 25 hours after the time of occurrence and
eleven telescopes observed the Parkes position over the following days
to months. Observations and results from each of these telescopes are
described in \S\ref{follow-up}.

\textbf{\textit{FRB 151230}} shows peak intensity near the centre of
our observing band, similar to some of the events described in
\citet{spitler2016repeating} for FRB 121102. The FRB is bright in the
upper 200 MHz of the band and disappears at the lower frequencies in
the band, below 1300 MHz. The burst is unresolved and shows scattering, 
possibly partly responsible for the non-detection at the lowest frequencies. 
We can determine the DM-index to be $\beta=2.00\pm0.03$.
This burst was discovered by the F-pipeline,
an alert was raised, and a trigger was issued to telescopes after an
hour of the detection. This burst was followed up by 12 telescopes
ranging from radio to gamma-ray wavelengths.

\textbf{\textit{FRB 160102}} is the highest-DM FRB yet observed with
DM = 2596.1$\pm$0.3 pc cm$^{-3}$, and has an inferred luminosity
distance of $17$~Gpc, assuming the nominal redshift $z = 2.1$ from the models of
\citet{ioka} and \citet{inoue} for the observed DM excess. We find indications 
of scattering and determine the DM-index to $\beta = 2.000\pm0.007$. 
For this FRB, a trigger was issued approximately one hour after the event and this burst was
followed up by 8 telescopes spanning radio to gamma-ray wavelengths.

\section{Follow-up studies}\label{follow-up}
\begin{figure*}
  \includegraphics[scale=0.4]{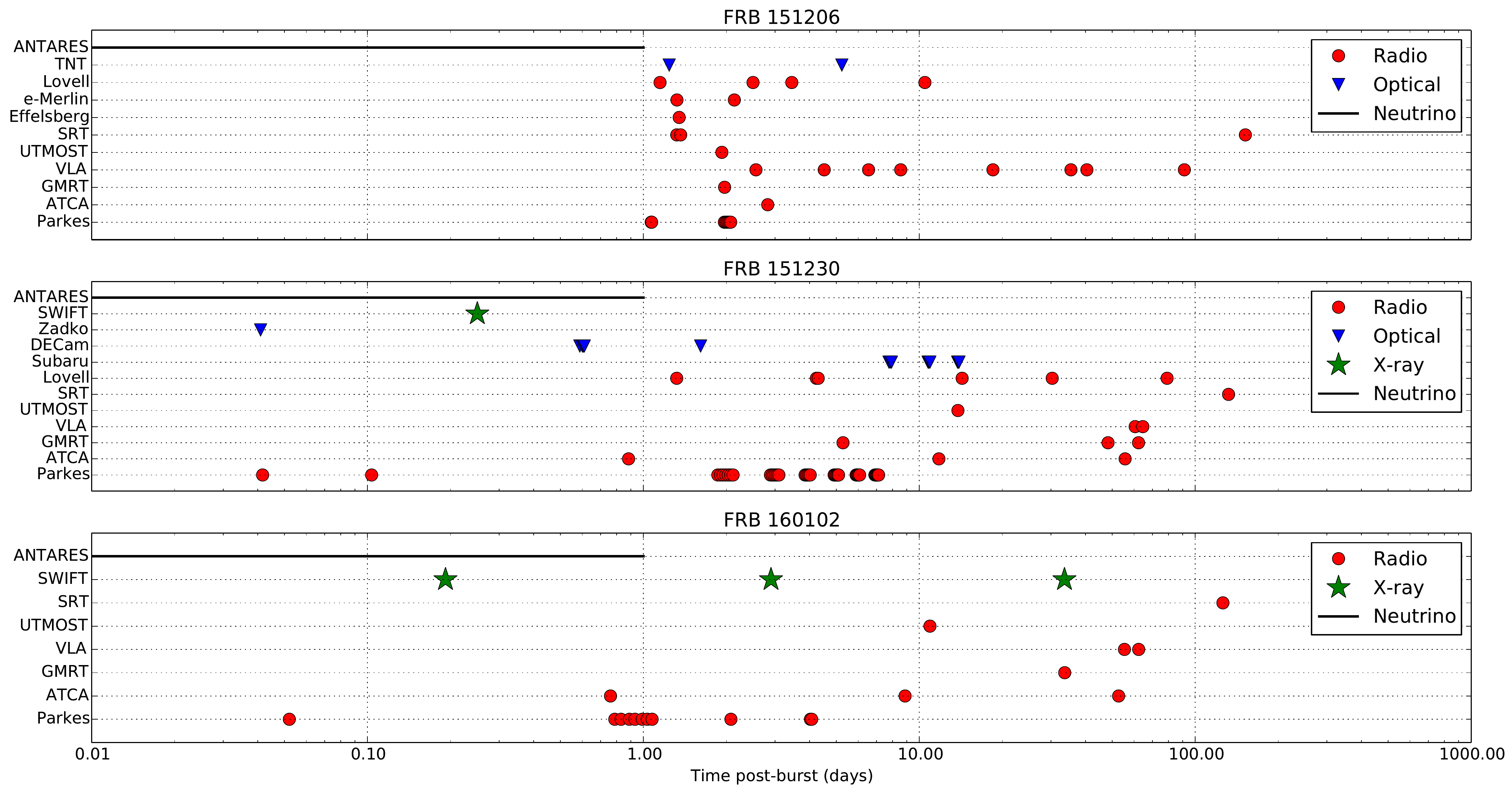} 
  \caption {Multi-messenger follow-up campaign for FRBs 151206,
    151230, 160102. The black line represents a part of the search for neutrino 
    counterparts with ANTARES over the window [T$_0-$1 day; T$_0$+1 day],
    where T$_0$ is the event time. No high energy follow-up was performed for FRB 151206 as
    it was Sun-constrained. Also, due to the delayed detection of FRB
    150610 the multi-messenger follow-up was restricted to an ANTARES search alone.}
  \label{figure:campaign}
\end{figure*}

Follow-up observations of each FRB's field were carried out with four
optical telescopes, nine radio telescopes, one high energy telescope and the
ANTARES neutrino detector.
Fig. \ref{figure:campaign} shows the summary of observations performed
on each field. Imaging observations with radio and optical telescopes
were performed in order to search for any variable or transient
sources that might be associated with the FRBs. Radio follow-up also
included searching for repeat pulses from each FRB location. A
complete record of all observations performed is included in
Tables~\ref{table:FRB1details} to \ref{table:FRB4details} in the
Appendix.
\subsection{Radio follow-up for repeat bursts}\label{followup_repeat}
Follow-up observations were performed with the Parkes telescope using the Berkeley Parkes Swinburne Recorder (BPSR)
observing setup \citep{BPSR} immediately after the discovery
of each real-time FRB. The Sardinia radio telescope
\citep[SRT;][]{Bolli} observed the FRB fields in single pulse search mode at a centre
frequency of 1548 MHz with a bandwidth of 512 MHz. Observations
were also performed by the Lovell and Effelsberg radio telescopes \citep{Lovell,Effelsberg} in
L-Band (1.4~GHz) and single pulse searches were performed with PRESTO
\citep{presto} around the DM of the FRB. The UTMOST telescope \citep{Bailes2017} also observed
three of the FRB fields (all except FRB 150610). The UTMOST 
observations were performed at 843 MHz with a bandwidth of 31 MHz in
fan beam mode with 352 fan beams covering $4\degree\times 2.8\degree$
 (see \citet{UTMOSTFRBs} for the details of this observing mode).  The
details of the time spent on each FRB field are listed in Table
\ref{table:repeats}. None of the observations showed repeated bursts
from their respective FRB fields.
\begin{table}
\caption{The time spent by the Parkes, SRT, Effelsberg, Lovell and
  UTMOST radio telescopes on the field of SUPERB FRBs to search for
  repeating pulses. None of the observations showed repeated bursts.}
\label{table:repeats}
\resizebox{8cm}{!}{
\centering
\begin{tabular}{|c|c|c|c|c|c|c|}
\hline
 FRB & Parkes & SRT & Effelsberg & Lovell & UTMOST  & Total\\
 &  T$_{\rm obs}$(hrs)   &  T$_{\rm obs}$(hrs)   &  T$_{\rm obs}$(hrs)  &  T$_{\rm obs}$(hrs)  &  T$_{\rm obs}$(hrs)  & (hrs)\\
\hline
\hline
FRB 150610 & 10  & - & - & - & - & 10 \\
FRB 151206 & 3 & 9.3 & 3 & 3.3 & 3.75 & 22.3\\
FRB 151230 & 36 & 2.9 & - &8.5 & 7.5 & 54.9\\
FRB 160102 & 9.2 & 2 &-&-& 4.7 & 15.9 \\
\hline
\end{tabular}}
\end{table}


\subsection {Radio interferometric follow-up for possible counterparts}\label{atca}

Radio imaging observations were performed using the Australian
Telescope Compact Array (ATCA) \citep{Wilson2011}, VLA, the Giant Metrewave Radio Telescope
(GMRT) \citep{GMRT} and the e-Merlin radio telescope \citep{e-Merlin}, spanning 4 to 8~GHz and 1 to 1.4~GHz. 
The details of the observations, data analysis and variability criteria are listed in
Appendix \ref{observation_detail}. Here we present the results of the
follow-ups and the implications of the variability are discussed in
\S\ref{end}.

\textit{\textbf{FRB 151206}}:
ATCA observed the field of FRB 151206 on 2015 December 9, 3 days after
the burst. Visibilities were integrated for 3 hours yielding a radio
map with an rms noise of $50~\upmu$Jy/beam at 5.5 GHz and 60
$\upmu$Jy/beam at 7.5 GHz. The declination of the FRB field ($\delta = -04^{\degree}$) was 
not favourable for ATCA observations, therefore no subsequent observations were 
performed and no variability analysis was conducted on these data.

\begin{figure*}
\centering
\subfigure[]{%
\includegraphics[scale=0.32]{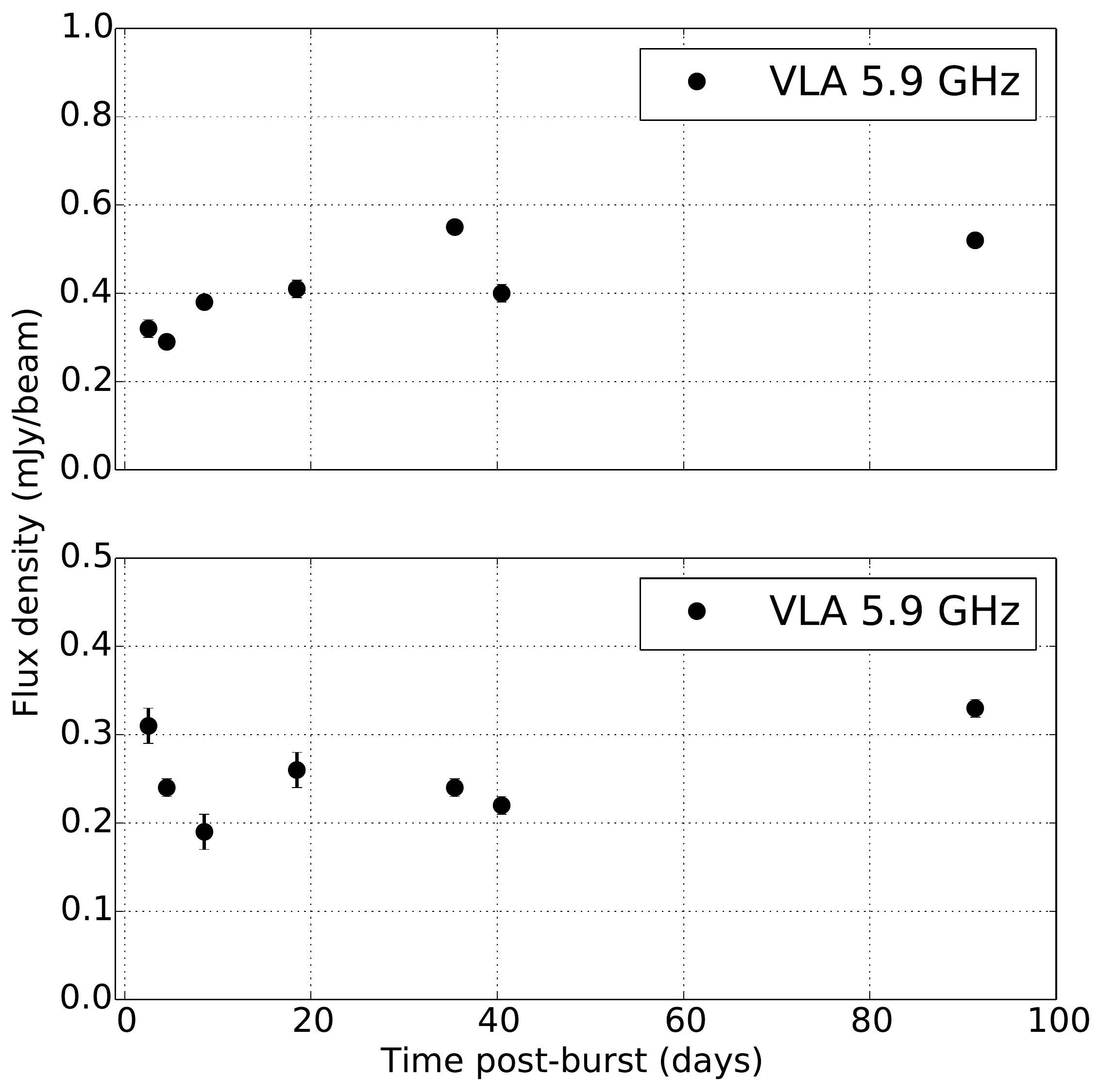} 
\label{figure:curves_VLA1}}
\quad
\subfigure[]{%
\includegraphics[scale=0.32]{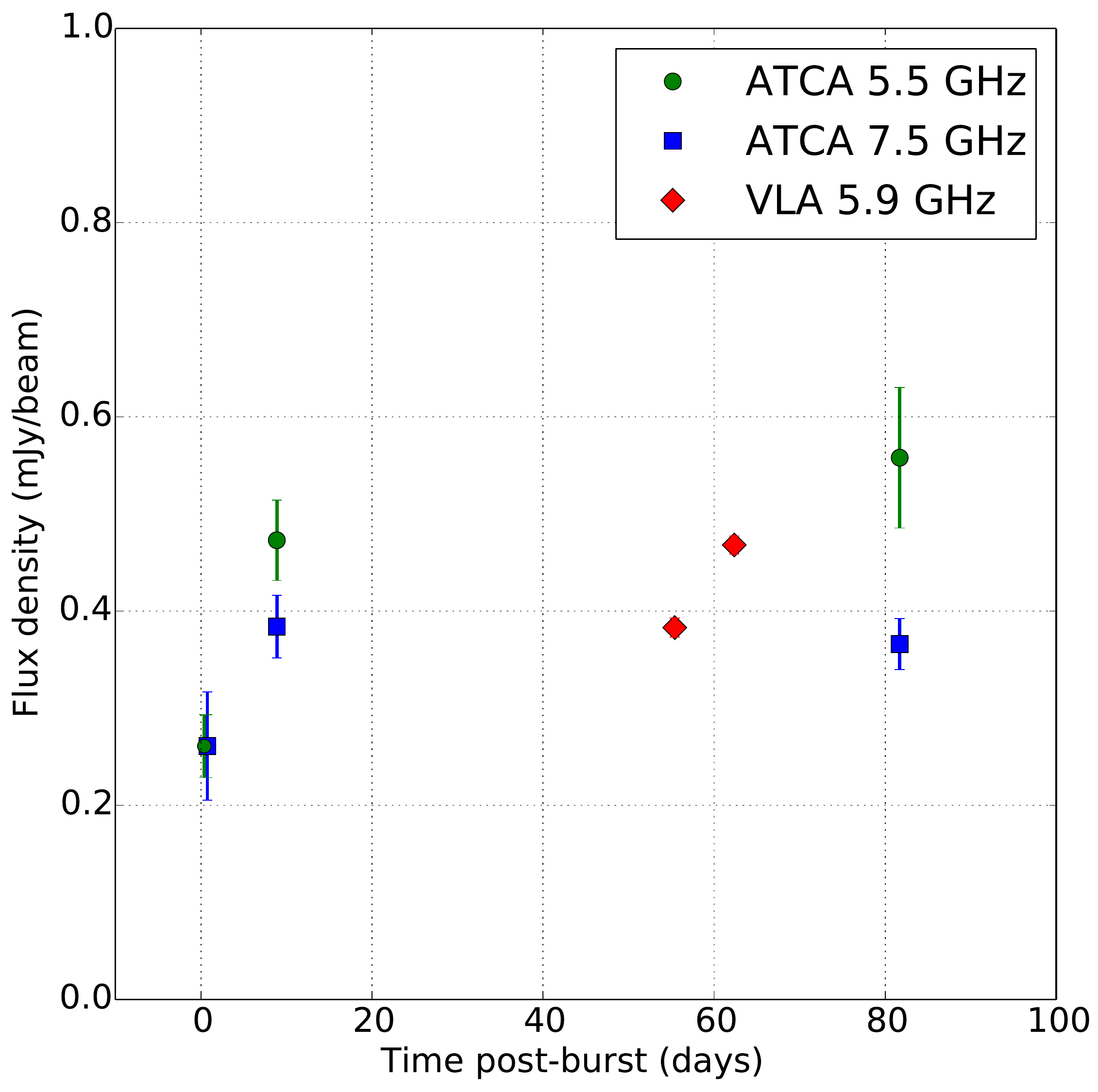} 
\label{figure:FRB3_ATCA}}
\caption{Left panel: The light-curves over 92 days of two sources in
  the field of FRB 151206 found to vary significantly in the VLA
  observations: 1921--0414 (top panel) and 1921--0412 (bottom
  panel). Right panel: The light curve of the significant variable
  source 2238--3011 in the field of FRB 160102. The fluxes and errors
  on fitting are derived from the task \textit{IMFIT} in
  \textit{miriad}. Note that the data have not been calibrated to the
  same absolute flux scale, and there may be systematic differences
  between different instruments. However, the data are self-consistent
  for variability analysis for each instrument.}
\end{figure*}

This field was observed for eight epochs with the VLA starting from 2015 December 8. 
The radio images reached an rms
of $10-25~\upmu$Jy/beam. Observations at epoch 3 were severely affected by RFI and hence
excluded from the analysis. To form mosaic images, each of the 7
single pointings were stitched together for every epoch and were
deconvolved using the {\sc clean} algorithm \citep{cleanalgo}. Two
significantly variable sources were detected in this field, details of which are listed in
Table \ref{table:table_variable}. Fig. \ref{figure:curves_VLA1} shows their light curves. No
non-radio counterpart was identified for either of the sources.

Observations were performed with the GMRT on 2015 December 9. The field
was observed for 4 hours and the map yielded an rms of 30
$\upmu$Jy/beam. No subsequent observations were 
performed and no variability analysis was conducted on these data.

The field was also observed with e-Merlin on 2015 December 7 and
8. Observations ran from 14:00--19:30 UTC on December 7 and
09:30--19:30 on December 8. A total of 1,945 overlapping fields were imaged and
then combined using the \textit{AIPS} task \texttt{FLATN}. The
combined image covered a circular area of 10$\arcmin$ diameter and has
an rms of 34 $\upmu$Jy/beam (beam size =171 $\times$ 31 mas,
PA=19.4$\degree$). At the declination of the source, snapshot imaging
is quite challenging for e-Merlin, so the combined full sensitivity image
from 1.5 runs was searched for significant detections with
\texttt{SExtractor} and nothing significant ($>$ 6-sigma) was found.

\textit{\textbf{FRB 151230}}: ATCA was triggered $\sim$1 day after the
event and visibilities were recorded for 8 hours. Subsequent
observations were performed on 2016 January 11 and 2016 February 24
for 9.5 and 4.5 hours, respectively. We performed a
variability analysis of all compact sources at 5.5~GHz and
7.5~GHz. Following the criteria described in Appendix
\ref{observation_detail}, we conclude that there are no significant
variable sources present in the field of FRB 151230.

Observations were performed using the VLA on 2016 February 29 and
2016 March 4 and images were produced at the center frequency of 5.9
GHz with an rms of $\sim$15 $\upmu$Jy/beam. All ATCA sources were
detected. None of the compact sources were found to be significantly variable.

GMRT observations were performed on 2016 January 6, 2016 February 17
and 2016 March 3. The integration times of 4 hours yielded an rms
of $\sim$30 $\upmu$Jy/beam at 1.4 GHz. None of the sources showed any
significant variability.

\begin{table*}
\caption{The results of the radio follow-up performed using the ATCA,
  VLA and GMRT on the fields of SUPERB FRBs. N$_{\rm total}$ denotes the
  total number of sources detected above 6-sigma and N$_{\rm analysis}$ are the number
  of sources used in the variability analysis. This excludes extended
  sources in the field of respective FRBs. N$_{\rm variable}$ denotes
  the number of significant variable sources detected in each field.}
\label{table:results}
\resizebox{18cm}{!}{ \centering
\begin{tabular}{|c|c|c|c|c|c|c|c|c|c|c|c|c|c|c|c|c|c|c|c|c|}
\hline
Telescope & \multicolumn{6}{|c|}{ATCA} &  \multicolumn{3}{|c|}{VLA} &  \multicolumn{3}{|c|}{GMRT}   \\
\hline
Centre freq. & \multicolumn{3}{|c|}{5.5 GHz} & \multicolumn{3}{|c|}{7.5 GHz} & \multicolumn{3}{|c|}{5.9 GHz}& \multicolumn{3}{|c|}{1.4 GHz} \\
\hline
 &  N$_{\rm total}$ & N$_{\rm analysis}$ &N$_{\rm variable}$&  N$_{\rm total}$ & N$_{\rm analysis}$ & N$_{\rm variable}$&N$_{\rm total}$ & N$_{\rm analysis}$& N$_{\rm variable}$&  N$_{\rm total}$ & N$_{\rm analysis}$ & N$_{\rm variable}$\\
\hline
\hline
FRB 151206 &  1 & -& -&  1 &  -&- & 10 & 10 & 2&13 & - & -\\
FRB 151230 & 9 & 6 & 0 & 5 & 2 & 0 & 25 & 20 & 0 & 27 & 18 & 0 \\
FRB 160102 & 12 & 10 & 1 & 12 & 10 & 0 & 21 & 19  & 0 & 48 & - & - \\
\hline
\end{tabular} }
\end{table*}

\textit{\textbf{FRB 160102}}: ATCA observed the FRB 160102 field on
2016 January 3, 2016 January 11 and 2016 February 24. The best map
yielded an rms of $\sim$40 $\upmu$Jy/beam at 5.5~GHz and $\sim$50
$\upmu$Jy/beam at 7.5~GHz.
The search for sources was performed over an area of sky that is twice the region of
the localisation error, i.e. a radius of 15$\arcmin$ because this FRB
was detected in the outer beam of the Parkes telescope.

The final variability analysis was performed on 10 compact
sources. Source 2238$-$3011 was found to vary significantly at 5.5~GHz
but not at 7.5~GHz. We identify it to the quasar 2QZ J223831.1$-$301152 from the ``Half a
Million Quasar Survey'' \citep{HMQ} at $z$ = 1.6. This source is also
present in the GALEX survey \citep{GALEX} (GALEX J223831.1$-$301152) and has a DSS \citep{DSS}
optical counterpart. Table \ref{table:table_variable} and Figure
\ref{figure:FRB3_ATCA} lists the details and light-curve of the source
2238$-$3011. The flux density of the source was observed to be rising
at ATCA epochs at 5.5~GHz.

The VLA observations were performed on 2016 February 26 and 2016
March 4. Flux densities were derived from mosaics with the best rms being
$\sim$10 $\upmu$Jy/beam. ATCA source 2238-3011 showed a low level variability with the fractional change 
(defined in Appendix \ref{observation_detail}), $\Delta$S $\sim$20$\%$
($<50\%$). None of the remaining sources were found to vary
significantly at 5.9~GHz.

The field was also observed with the GMRT on 2016 February 6. The
integration of 4 hours yielded an rms of $\sim$30
$\upmu$Jy/beam. This GMRT epoch was used to cross-check sources detected in the
ATCA and VLA images and no variability analysis was performed on these data.

\begin{table*}
\caption{Radio variable sources in the field of FRB 151206 and FRB
  160102. The errors in RA and DEC are in arcseconds and
  are presented in brackets. Columns 4 and 5 list $\chi^{2}$ and
  $\chi^{2}_{\rm thresh}$ values. The $\chi^{2}_{\rm thresh}$ values
  are upper-tail critical values of chi-square distribution with $N-1$
  degrees of freedom. Columns 6 and 7 list $m_{\rm d}$, and
  $\Delta$S values. These variability indices are defined
  in Appendix \ref{observation_detail}.}
\label{table:table_variable}
\centering
\begin{tabular}{|c|c|c|c|c|c|c|}
\hline
 Name & RA & DEC & $\chi^{2}$ &  $\chi^{2}_{\rm thresh}$& $m_{\rm d}$ & $\Delta$S   \\
 & & & & & ($\%$) & ($\%$) \\
\hline
\hline
& & FRB 151206 field & & & \\
\\
VLA1921-0414 & 19:21:27.21 (0.2)  & $-$04:14:55.67 (0.2) & 478.6 & 24.3& 21.3 & 63.4 \\
VLA1921-0412 & 19:21:43.85 (0.2) & $-$04:12:17.43 (0.2) & 91.0 & 24.3&  16.8 & 54.7 \\
\hline
& & FRB 160102 field & & & \\
\\
ATCA2238-3011 & 22:38:31.17 (0.2) & $-$30:11:51.38 (0.6) & 24.16 & 13.8& 26.4 & 69.0 \\
\hline
\end{tabular}
\end{table*}

The results of the radio follow-up are summarised in Table
\ref{table:results}.


\subsection{Follow-up at non-radio frequencies}
We have carried out optical and high energy follow-up and searched for
neutrino counterparts to these four SUPERB FRBs. The results are presented in
this section and the details of the observations and magnitude limits
are listed in Appendix \ref{other}.

\subsubsection{Thai National Telescope (FRB 151206)} 
The observations were performed with ULTRASPEC on the Thai National Telescope (TNT) 
on the night of 2015 December 7. Four optically variable sources were found in the field of FRB 151206.
The change in magnitude $\Delta$mag provides a measurement of the variability of a given
source in the field, such that $\Delta$mag > 0 reflects a dimming
source. The only source detected with a negative $\Delta$mag is also bright
at infrared wavelengths, with $J = 9.38$, $H = 8.31$, $K = 7.93$ respectively from 2MASS \citep{2MASS}. Further photometric
observations of the four variable sources were obtained using the
0.5-m robotic telescope ``pt5m'' \citep{hardy}. In all cases, the
variability seen for these sources can be explained by stellar
variability, either eclipsing, ellipsoidal or stochastic (accretion,
flaring etc).

\subsubsection{Subaru Telescope (FRB 151230)}
We performed follow-up imaging observations of the field of FRB
151230 in the $g$-, $r$-, $i$-bands on 2016 January 7, 10 and 13,
with Subaru/Hyper Suprime-Cam that covers a 1.5 deg diameter
field-of-view. The images taken on January 13 were used as the
reference images and were subtracted from the images of January 7
and 10 using the HSC pipeline \citep{Bosch2017}. Ninty-seven
variable source candidates with either positive or negative flux
difference were detected in the error circle
of FRB 151230 on the differential images. These candidates were
examined by eye, and approximately half of them appear to be real objects while
the other half are artefacts by subtraction failure. Most of the
real variable sources are likely to be either Galactic variable stars (point sources
without host galaxy) or AGNs (variable sources located at
centres of galaxies). There are three objects associated with
galaxies and offset from galaxy centres, which are most likely
supernovae. This number is consistent with those detected outside
the FRB error region considering the area difference, and also
consistent with a theoretically predicted number of supernovae
with the depth and cadence of our observations \citep{Niino}. No
object shows evidence for an association with
the FRB, although we cannot exclude the possibility that one of
them is associated. The nature of the variable objects
will be investigated and discussed in detail in a forthcoming
paper (Tominaga et al. in prep).
\subsubsection{DECam (FRB 151230)}
We obtained DECam {\it u-g-r-i} dithered images centred on the
coordinates of FRB 151230, with observations taken approximately 14 hours after the detection at
Parkes. The field was also re-observed with the $g$ filter $\sim$\,39
hours after the FRB detection. We searched these $g$-band images for
transient sources ($>$10-sigma significance) between the two consecutive nights, 
within the localisation error region of 15$\arcmin$, using {\it Mary} pipeline (Andreoni et al.,under review).
We detected 5 variable sources and 4 of them were
cataloged\footnote{NASA/JPL SB identification system:
  ssd.jpl.nasa.gov} as small bodies, i.e. Main-belt asteroids. A fifth
object was detected at a magnitude $g$\,=\,22.51$\pm$0.08 on 2016 January 1,
which had not been detected on the previous night, 2015 December 31 ($g
<$\,23.37 at 5-sigma confidence). This transient is located at
RA=9:40:56.34, DEC=$-$3:27:38.29 (J2000) and is not present in the
NASA/JPL small body catalog but is most likely to be an asteroid unrelated to FRB. However, it is not detected in the {\it
  u-g-r-i} images taken on 2015 December 31. All other transient events
were rejected as bonafide transients due to poor local subtraction and
bad pixels after a visual inspection of the residuals.

We have also compared the radio sources detected in the GMRT and ATCA
images with the DECam images to look for optical counterparts; more
details are given in Appendix \ref{other}.

\begin{figure*}
  \includegraphics[scale=0.7]{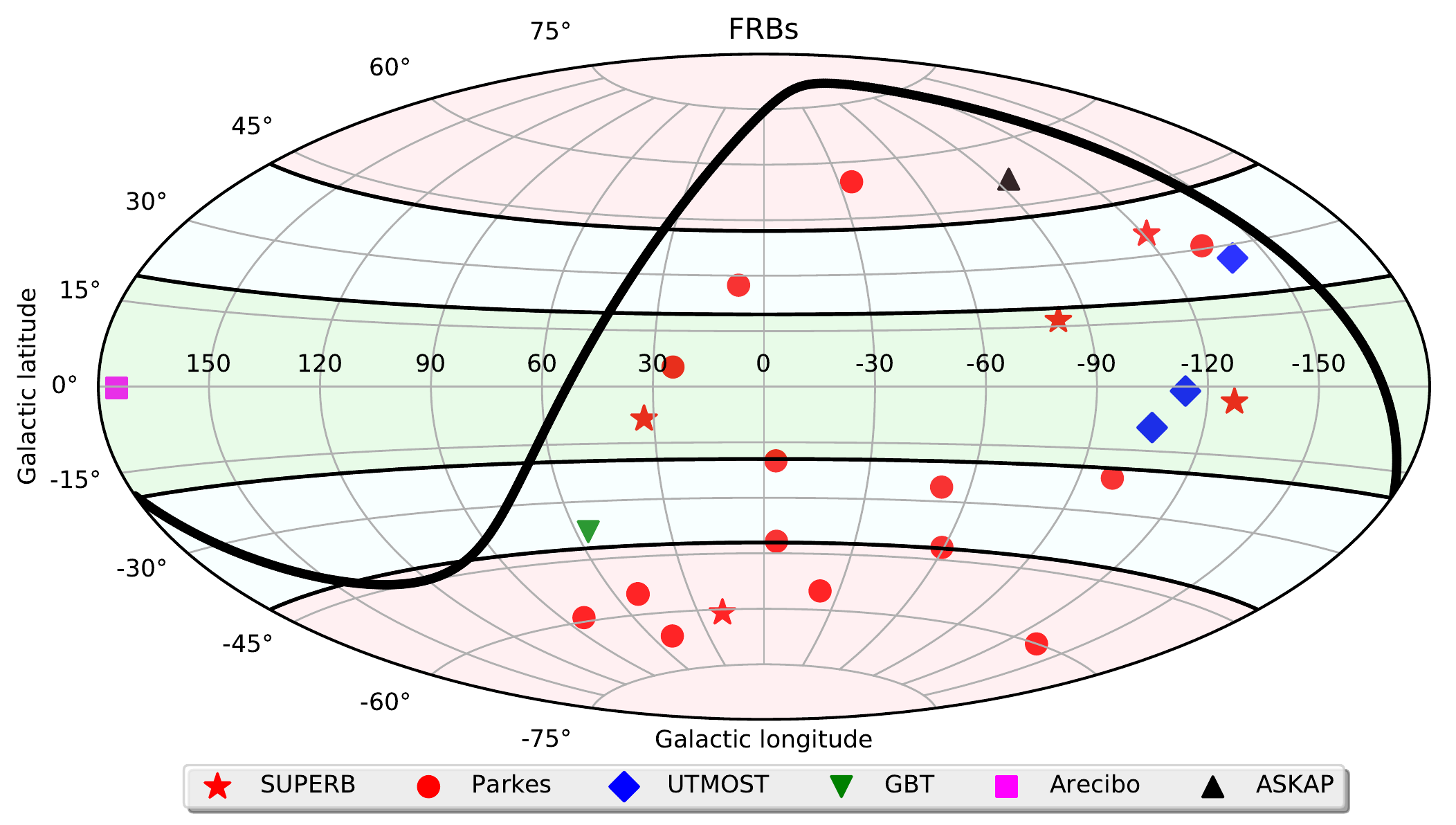}
  \caption {An Aitoff projection of the sky distribution of all published FRBs. The
    shaded regions show the three Galactic latitude bins in Table
    \ref{table:time_number}. The bold black line shows the horizon
    limit of the Parkes radio telescope.}
  \label{figure:FRBdist}
\end{figure*}

\begin{table*}
  \caption{Time on sky in the three latitude bins for recent surveys
    conducted at the Parkes telescope: the High Time Resolution
    Universe survey \citep[HTRU;][]{BPSR}, observations of rotating
    radio transients, FRB follow-up, the SUPERB survey, and
    observations of young pulsars for \textit{Fermi} timing.  All
    surveys made, or make, use of the multi-beam receiver and have
    equivalent field of view and sensitivity limits. The FRB sky rates for respective latitude bins 
    are quoted with 95$\%$ confidence.}
\label{table:time_number}
\begin{tabular}{|c|c|c|c|c|c|c|c|c|c|c|}
  \hline 
  Galactic latitude & HTRU & HTRU & RRAT  & FRB & SUPERB & Fermi & Misc & Total & $N_{\mathrm{FRBs}}$ & $R_{\mathrm{FRB}}$  \\
       $ |b|$                       & medlat & hilat & search & follow-up & & timing & time  &  time &  &  \\
  (deg)      & (hrs)     & (hrs) & (hrs)    & (hrs)         & (hrs) & (hrs) & (hrs) &  (hrs) & & FRBs sky$^{-1}$ day$^{-1}$ \\
  \hline
  \hline
  $|b| \leq19.5\degree$ & 1157 & 402 & 483 & 0 & 700 & 281 & 0 & 3024 & 4 & 2.4$^{+ 3.1}_{-1.5}$ $\times$ 10$^{3}$\\
  $19.5\degree < |b| < 42\degree$ & 0 & 942 & 28 & 50 & 1115 & 10 & 100 & 2245 & 6 &  4.8$^{+ 4.6}_{-2.7}$ $\times$ 10$^{3}$\\
  $42\degree \leq |b| \leq 90\degree$ & 0 & 982 & 39 & 60 & 907 & 9 & 90& 2088 & 9 &7.8 $^{+ 5.8}_{-3.7}$ $\times$ 10$^{3}$\\
  \hline
\end{tabular} 
\end{table*}

\subsubsection{The Zadko Telescope (FRB 151230)}
On 2015 December 30, the Zadko telescope was shadowing the Parkes
telescope at the time of the discovery of FRB 151230. However, due to
technical difficulties, the first science images were taken at
18:03:20.6 UTC, i.e $\sim$1 hour after the FRB event. Following this
initial imaging, a series of 19 images of 5 tiles each were obtained
during about 2 hrs through to the end of the night. Each image had an
exposure time of 60 seconds in the $r$-band. The localisation error
region (15$\arcmin$) around FRB 151230 is completely covered by the
central image of the tiles and partly contained ($\sim 33\%$) in the
peripheral images.

We analysed the individual images to search for new optical or
variable sources in the field of FRB 151230. We particularly focused
on the central image of the tile that fully covers the error radius
around the FRB position. We found no convincing new or variable
optical sources.

\subsubsection{High energy follow-up (FRB 151230, FRB 160102)}

We acquired follow-up observations with \textit{Swift} on FRB 151230
burst on 2015 December 30 at 23:14:45 UTC, about 7 hours after the FRB
for a duration of 2.05 ks. No sources were detected above a 2.5-sigma limit in the X-ray
image. The data were analysed using the tools available at the Swift
website \citep{,SwiftOnline, Evans2009} on an
observation-by-observation basis. Count rates were converted to X-ray
flux assuming a GRB-like spectral index of $-$2.0 and Galactic
H\textsc{I} column density estimates from the HEAsoft tool ``nH''.

We acquired 3 epochs on the field of FRB 160102 with the Swift XRT of
durations 3.5 ks, 3.3 ks, and 1.8 ks, respectively. No sources were
detected above a 2.5-sigma limit in any of the images. We did not
trigger \textit{Swift} for FRB 150610 (due to the delay in its
detection), nor FRB 151206 (as it was Sun constrained for 31 days
after the FRB).

No Swift-BAT REALTIME triggers were issued for short duration gamma-ray transients 
during the follow-up observations for each FRB field.

\subsubsection{ANTARES follow-up (all FRBs)}
Multi-messenger observations with high-energy neutrino telescopes can help to constrain the FRB origin 
and offer a unique way to address the nature of the accelerated particles in FRBs. The ANTARES telescope
 \citep{Ageron11} is a deep-sea Cherenkov neutrino detector, located 40 km off Toulon, France, in the 
 Mediterranean Sea and dedicated to the observation of neutrinos with ${E_\nu\gtrsim100~\rm GeV}$. ANTARES 
 aims primarily at the detection of neutrino-produced muons that induce Cherenkov light in the detector. Therefore, 
 by design, ANTARES mainly observes the Southern sky (2$\pi$ steradian at any time) with a high duty cycle. 
 Searches for neutrino signals from the four detected FRBs have been performed within two different time windows 
 around the respective FRB trigger time, T$_0$, within a 2$^\circ$ radius region of interest (ROI) around the 
 FRB position (3-sigma ANTARES point spread function for the online track reconstruction method). The first 
 time window $\rm{\Delta T_1}$ = [T$_0-$500 s; T$_0$+500 s] is short and was defined for the case where FRBs 
 are associated with short transient events, e.g. short Gamma-Ray Bursts \citep{Baret11}. A longer time 
 window $\rm{\Delta T_2}$ = [T$_0-$1 day; T$_0$+1 day] is then used to take into account longer delays 
 between the neutrino and the radio emission. The number of atmospheric background events within the 
 ROI is directly estimated from the data measured in the visible Southern sky using a time 
 window $\rm{\Delta T_{back}}$ =  [T$_0-$12 hr; T$_0$+12 hr]. The stability of the counting rates 
 has been verified by looking at the event rates detected in time slices of 2 hours within 
 $\rm{\Delta T_{back}}$. Within $\rm{\Delta T_1}$ and $\rm{\Delta T_2}$, no neutrino events were found
  in correlation with FRB 150610, FRB 151206, FRB 151230 or FRB 160102.

\section {Results and Discussion } \label{end}
\subsection{Cosmological implications of high DM FRBs}

Assuming FRBs are extragalactic, the DM may be divided into contributions along the line-of-sight from the ISM in the Milky Way (DM$_{\rm Galaxy}$), the 
Intergalactic Medium (DM$_{\rm IGM}$), a host galaxy (DM$_{\rm host}$) and the circum-burst medium (DM$_{\rm source}$):
\begin{equation}
\rm DM_{\rm FRB} = DM_{\rm Galaxy} + DM_{\rm IGM} + DM_{\rm host} + DM_{\rm source}.
\end{equation}
For all the FRBs reported here, the DM$_{\rm Galaxy}$ contribution is minor ($<10\%$ of the total observed DM). It is currently difficult to disentangle the DM contributions of the remaining DM terms for these bursts. 
\citet{Xu2015} showed the DM$_{\rm host}$ to peak in the range of 30 to 300~pc~cm$^{-3}$ for different inclination angles of a spiral galaxy and average DM$_{\rm host}$ to be 45~pc~cm$^{-3}$ and 37~pc~cm$^{-3}$ for a dwarf and an elliptical galaxy respectively. In such cases, the
remaining DM is expected to arise from the IGM if the sources are cosmological in nature.

If the DM of our FRBs is indeed dominated by the IGM contribution, then 
we are potentially probing the IGM at redshifts beyond $z \ga 2$.  If we can find FRBs with DM $\ga 3000$~pc~cm$^{-3}$, 
we could begin to probe the era in which the second helium reionisation
in the Universe occurred \citep{Fialkov2016}, which is important
for determining the total optical depth to reionisation of the Cosmic Microwave Background (CMB), $\tau_{\rm CMB}$.
We note that we discovered FRB 160102 soon after our pipelines were modified to allow for
DM searches above 2000~pc~cm$^{-3}$ (the current upper limit is 10,000~pc~cm$^{-3}$).
Even in the absence of scattering being a dominant factor higher sensitivity instruments will likely be needed to probe such high redshifts. 

\subsection{FRB latitude dependence revisited} \label{update}
With an ever increasing sample of FRBs detected with the BPSR backend
it is worthwhile to revisit the Galactic latitude dependence in FRB
detectability first examined in
\citet{Petroff2014a}. Table~\ref{table:time_number} summarises the
data from SUPERB, as well as several other projects using BPSR that have each
observed the sky with essentially the same sensitivity to FRBs resulting in the total of 19 bursts. We
consider three 
regions on the sky, delineated in Galactic
latitude as follows: $|b| \leq 19.5\degree$, $19.5\degree < |b| <
42\degree$, and $42\degree \leq |b|$. The time on sky in each of these
regions and the updated FRB rate at the 95$\%$ confidence level are presented in the
table. Fig.~\ref{figure:FRBdist} shows these FRBs on an Aitoff
projection in the Galactic coordinate frame. For the studies
considered here Parkes has spent $\sim42\%$ of the total time in the
lowest Galactic latitude region (this is mostly driven by pulsar
searches and/or continued monitoring studies). Despite this only 4 of
the 19 bursts have been found in this range. At the highest latitudes
9 FRBs have been detected in $\sim 40\%$ of the total time. 
\begin{figure}
 \includegraphics[scale=0.3]{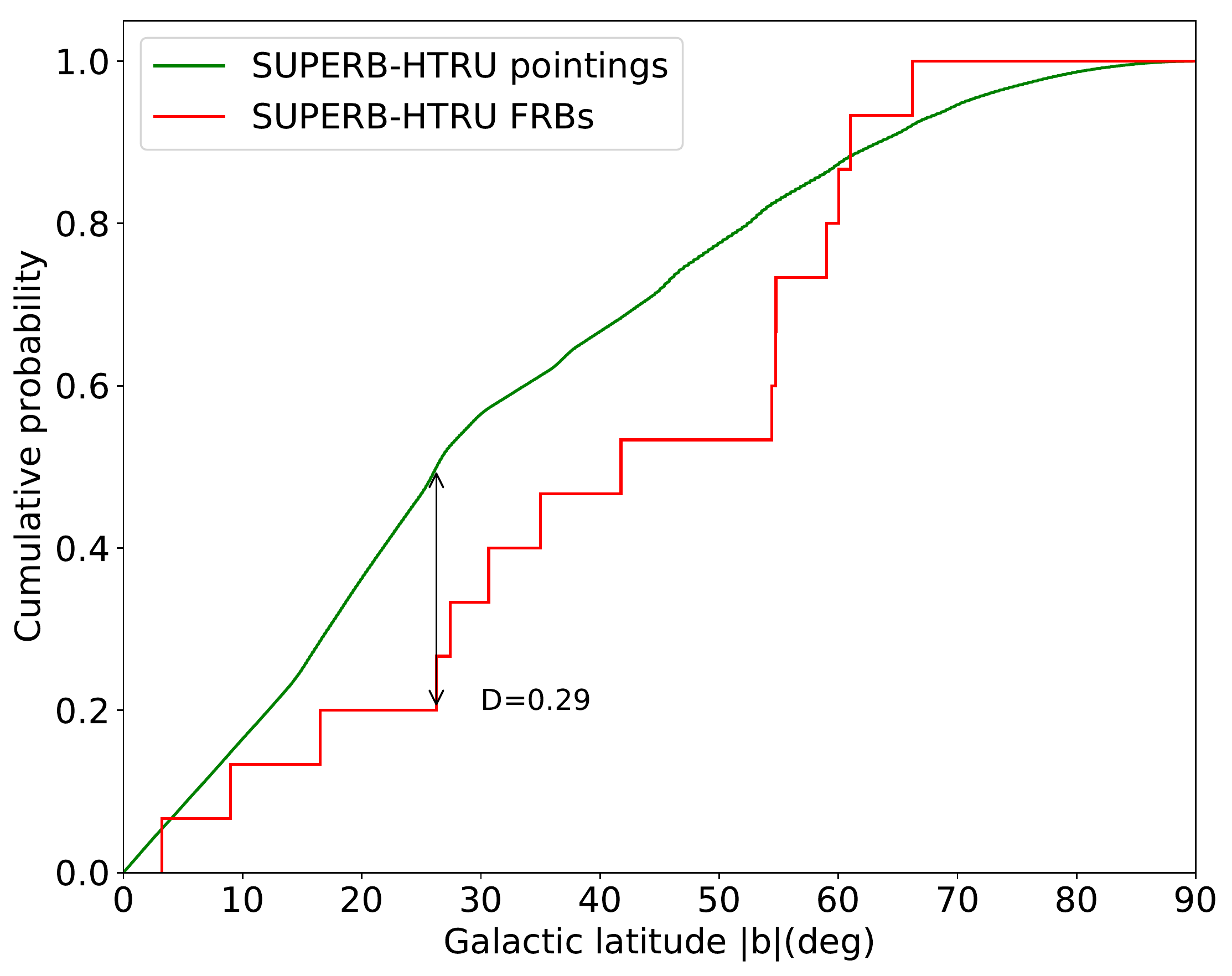}
 \caption {The observed cumulative distribution of Galactic latitude $|b|$ of FRBs detected in HTRU and SUPERB
  and the expected integration-time-weighted cumulative distribution of Galactic latitude $|b|$ for isotropically distributed FRBs.
 A K-S test indicates that the FRB distribution does not deviate significantly from isotropy.}
 \label{figure:KS_test}
\end{figure}
We performed a Kolmogorov-Smirnov test (K-S) between the expected cumulative distribution of $|b|$ for isotropically distributed 
FRBs based on the integration-time-weighted Galactic latitudes of the 
combined HTRU-SUPERB survey pointings, 
and the observed cumulative distribution of the 15 FRBs (see Fig \ref{figure:KS_test}). 
We obtain the KS statistic $D$ and $p$ values of 0.29, 0.10 respectively, and conclude that departure from isotropy is not significant.
Thus any disparity in the FRB rate with Galactic latitude has low significance ($< 2\sigma$) in our now larger sample of 15 FRBs. If such a disparity exists, it could be explained by diffractive scintillation boosting at high Galactic latitudes as discussed in \citet{JP}.

\begin{figure*}
\centering
\subfigure[]{%
\includegraphics[scale=0.4]{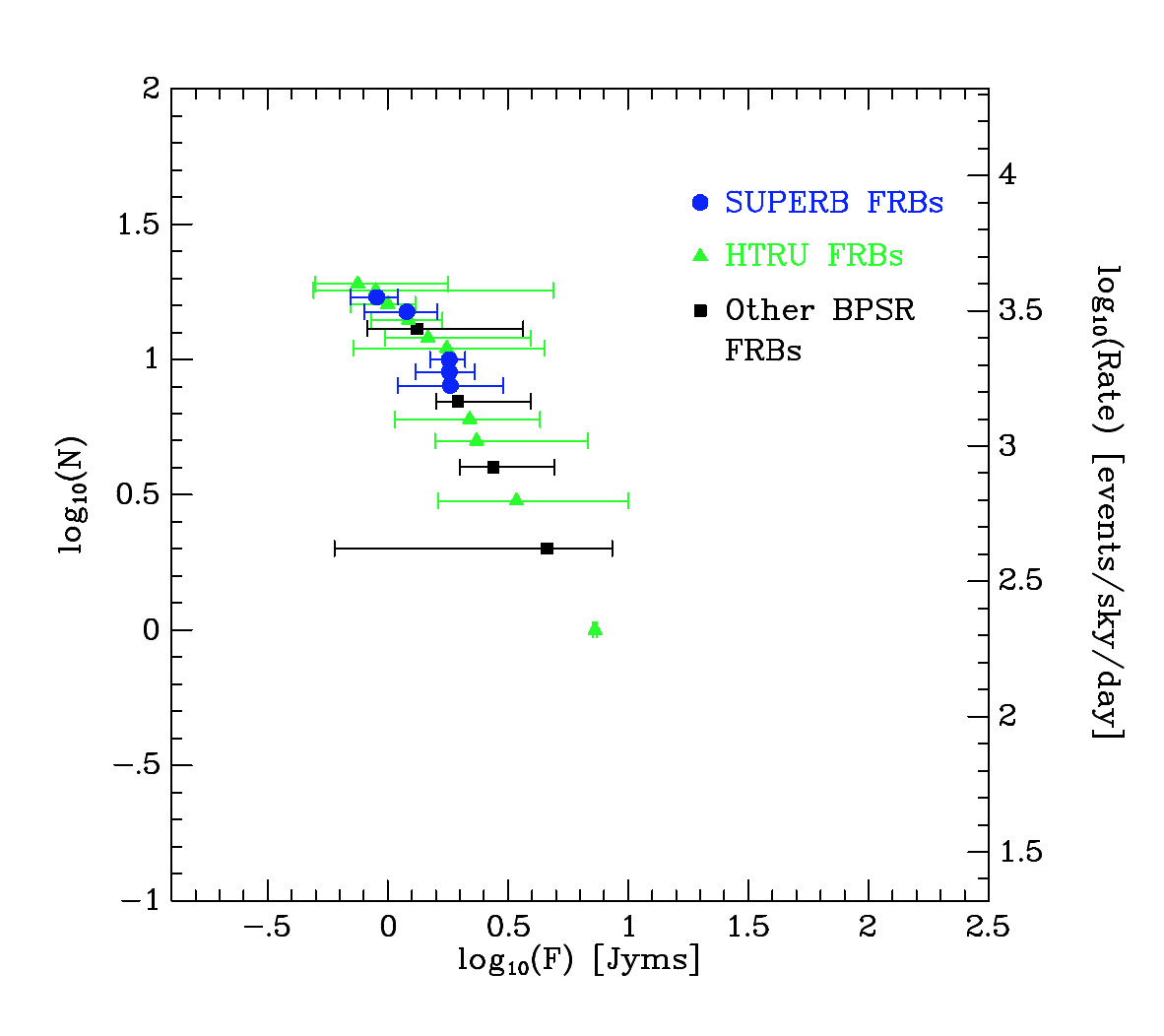} 
\label{source_counts}}
\quad
\subfigure[]{%
\includegraphics[scale=0.4]{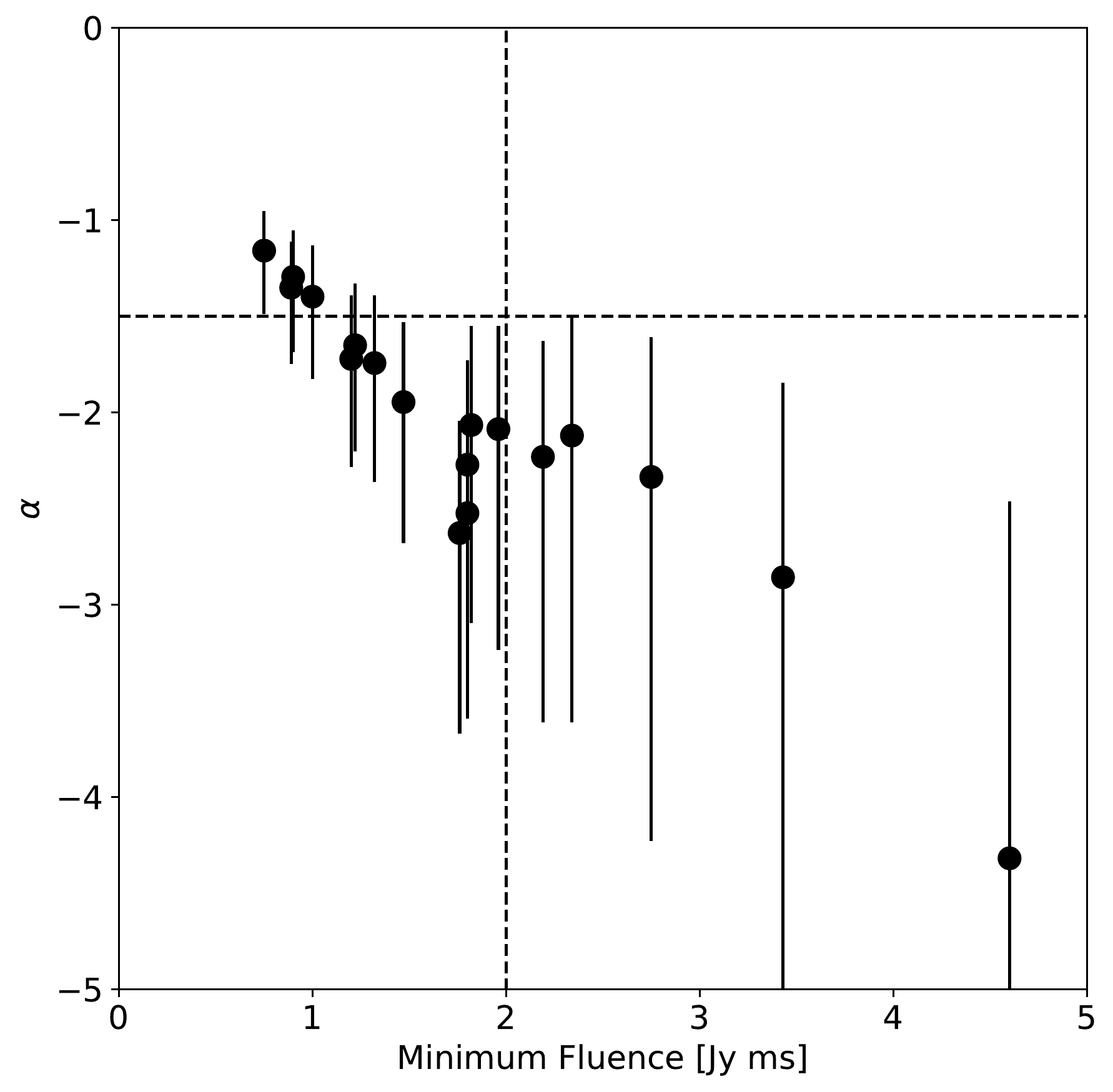} 
\label{figure:JP}}
\caption{Left panel: The source count distribution of Parkes FRBs. The sky rate is
   indicated on the right, normalised to rate of $1.7 \times 10^3$ FRBs 
   sky$^{-1}$ day$^{-1}$ for $\mathcal{F} > 2$~Jy ms (see \S\ref{skyrate}).
  Right panel: The slope $\alpha$ of the integral source counts obtained using the maximum likelihood method \citep{Crawford1970}. We obtain
  a slope of $\alpha = -2.2_{-1.2}^{+0.6}$ for FRBs above a fluence completeness limit of 2~Jy~ms in our updated sample of 19 FRBs. The vertical dashed line indicates the fluence completeness limit and the horizontal 
  dashed line indicates $\alpha = -3/2$, the slope expected for constant space density sources distributed in a Euclidean Universe.}
\end{figure*}
\subsection{FRB populations and distributions}\label{logNlogF}
Sources with constant space density in a Euclidean Universe yield an
integral source counts, $N$, as a function of fluence,
$\mathcal{F}$, the so-called ``logN-log$\mathcal{F}$''-relation, with a slope of
$-3/2$. The relation flattens in $\Lambda$CDM cosmologies, depending
on the redshift distribution of the sources being probed, and depends
to some extent on the luminosity function of the sources, and
observational factors like the effects that DM smearing have on the
S/N of events \citep{simu,Harish,chime}.

In Fig. \ref{source_counts}, we present the FRB source count
distribution as a function of fluence, for FRBs found with the BPSR
instrument at Parkes. The sample consists of 10 FRBs found in the HTRU
survey \citep[][]{Thornton,Champion}; Petroff et al (in prep), 5 FRBs found with SUPERB \citep[][FRB 150418]{keane2016host}
and this paper, and 4 FRBs found at Parkes with the
same instrumentation and search technique \citep[][]{Vikram, RaviScience, FRB140514,
FRB150215}. 

\begin{figure}
 \includegraphics[scale=0.45]{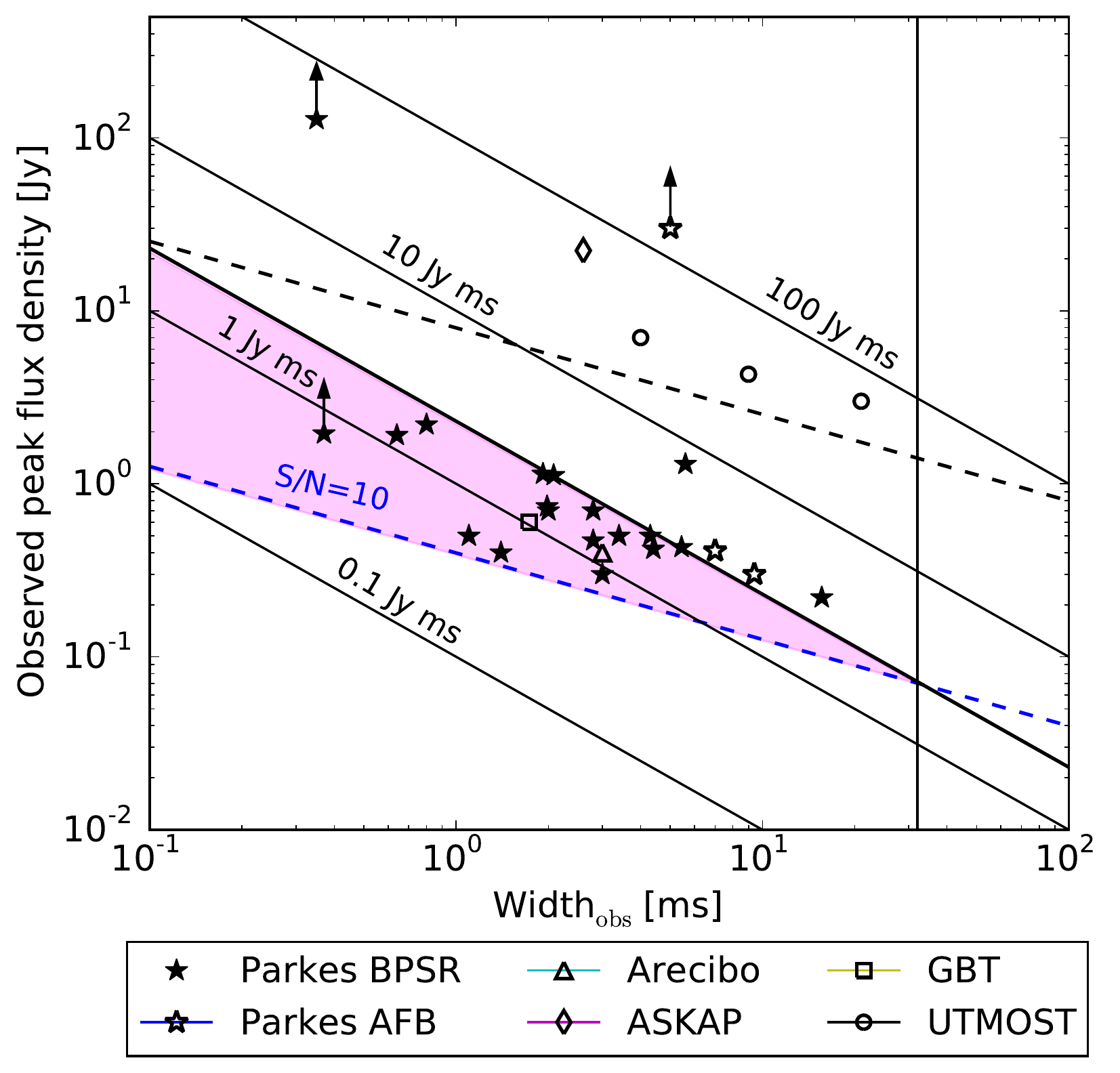}
 \caption {The observed peak flux density and observed
   width for all known FRBs. The sensitivity limits and fluence
   completeness region for BPSR Parkes events are indicated. These do
   not apply to other events which are shown for reference only.}
 \label{figure:FRBfluence}
\end{figure}

We note the following caveats about the logN-log$\mathcal{F}$ distribution. Firstly, the fluences are lower 
limits, as most of the FRBs are poorly localised within the Parkes beam pattern. Secondly, all FRB surveys 
are incomplete below some fluence, due to the effects of DM smearing, scattering and the underlying width distribution of the 
events (see \S\ref{skyrate} and Fig. \ref{figure:FRBfluence}). Although both these affect the shape of 
logN-log$\mathcal{F}$, simulations performed by \citet{simu} show that the slope of the relation is 
mainly set by cosmological effects. They found $\alpha = -0.9 \pm 0.3$ for the 9 HTRU FRBs. 

We measure a slope of the integral source counts using the maximum likelihood method \citep{Crawford1970} and obtain 
$\alpha~=~-2.2_{-1.2}^{+0.6}$ for FRBs above a fluence limit of 2~Jy~ms as shown in Fig \ref{figure:JP}. This is consistent with 
the source count slope for Parkes FRBs
found by \citet{Macquart2017}, who find $\alpha = -2.6_{-1.3}^{+0.7}$ 
The large uncertainty in $\alpha$ is due to the small sample size. Similarly
to \citet{Macquart2017}, we are unable to rule out that the source counts
are not Euclidean ($\alpha = -3/2$).

\begin{figure*}
\centering
\subfigure[]{%
\includegraphics[scale=0.35]{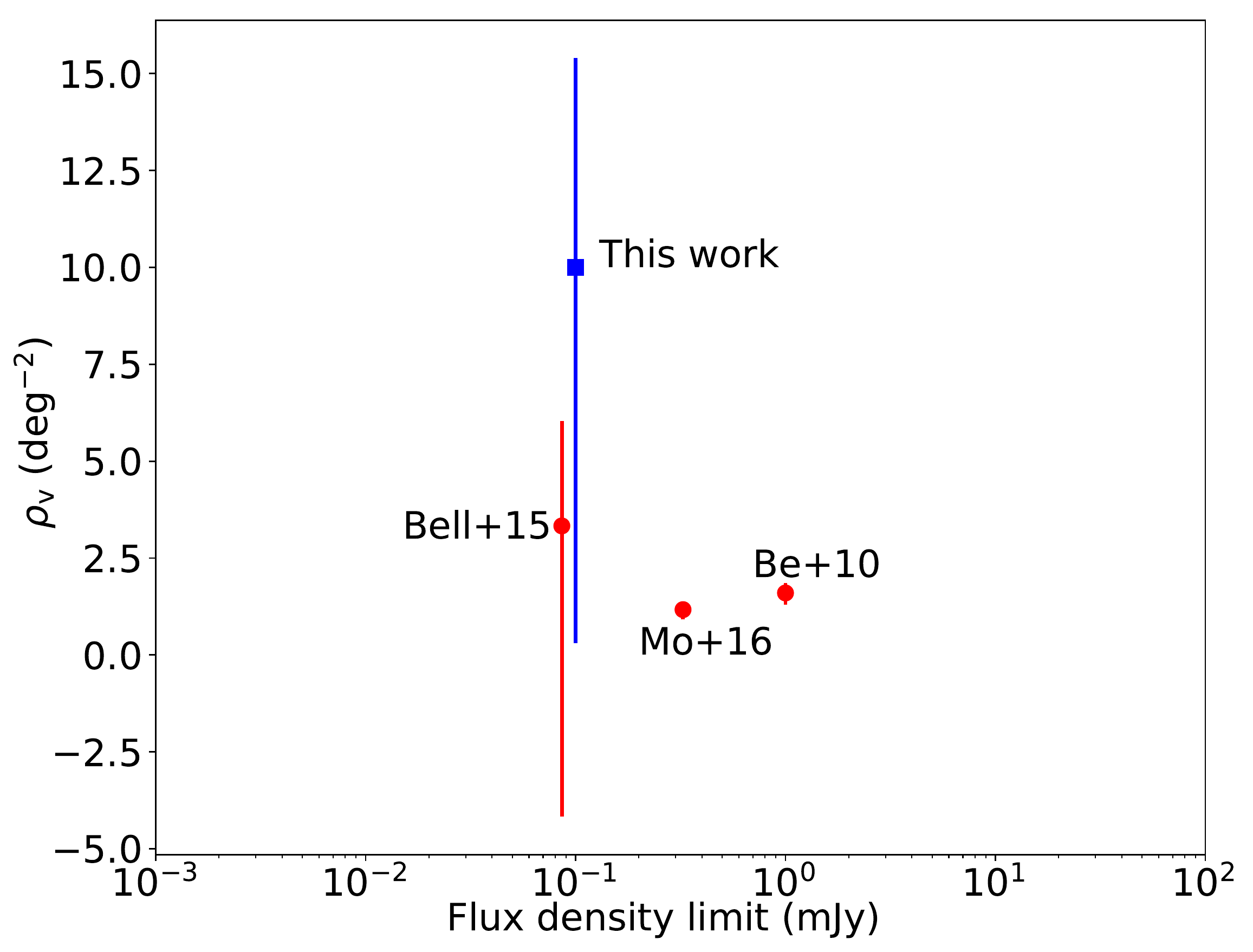} 
\label{figure:source_density}}
\quad
\subfigure[]{%
\includegraphics[scale=0.35]{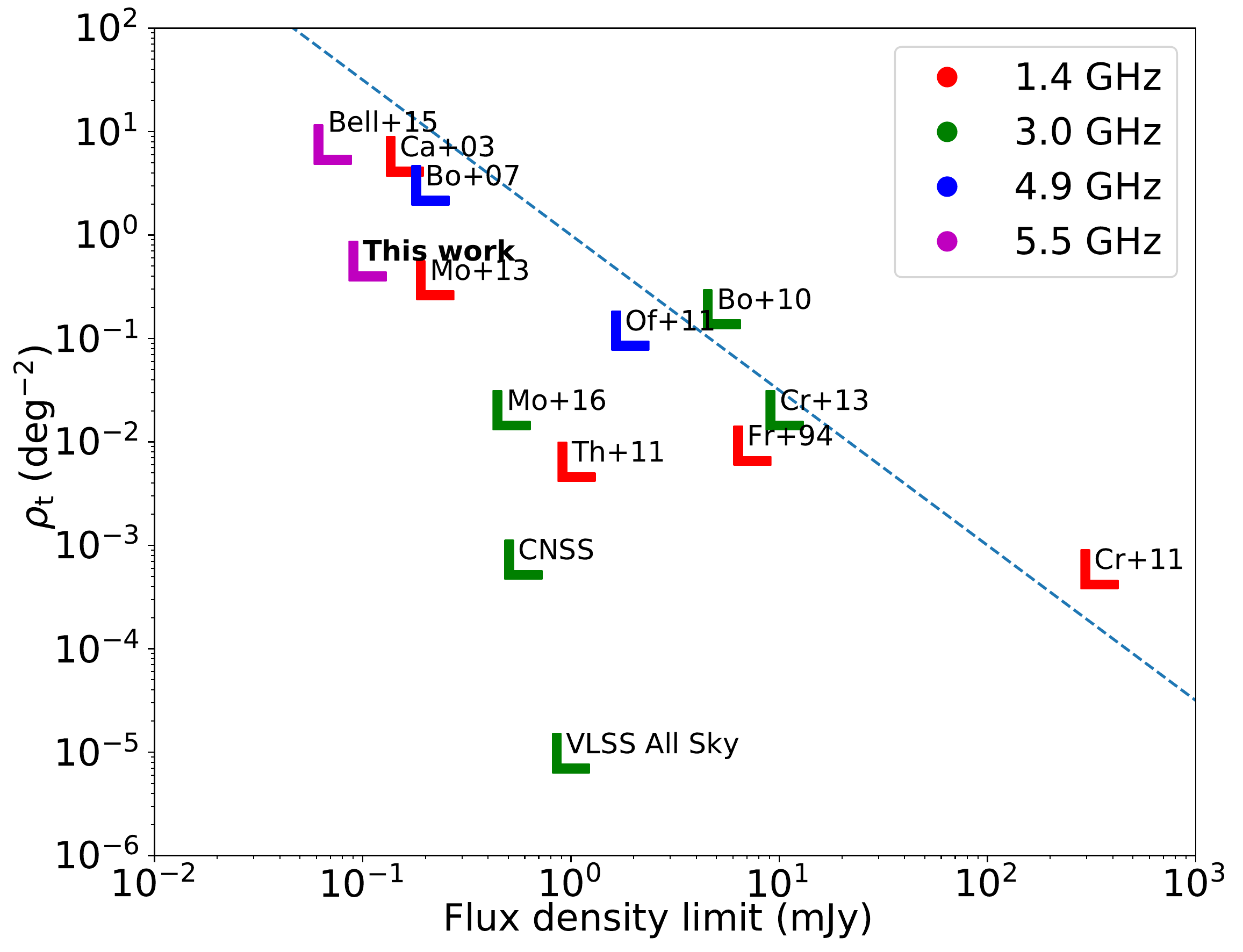} 
\label{figure:transient_density}}
\caption{Left panel: The density of significantly variable radio
  sources as a function of flux density in surveys made at $\sim$5 GHz
  and 3 GHz by \citet{bell}, this work, \citet{becker} and
  \citet{mooley}. The density of significantly variable sources is
  consistent within a 1-sigma Poisson error for surveys done in the
  past. Right panel: The density of transient radio sources in surveys
  conducted at 1.4~GHz \citep{Carilli2003,Mooley2013,Thy2011,Frail1994,Croft2011}, 3~GHz \citep[][CNSS pilot, CNSS, VLSS]{Bower2010,Croft2013,mooley}, 
 4.9~GHz \citep{Bower2007,ofek}, and 5.5~GHz (This work, \citet{bell}) as a function of flux density. 
   The dashed blue line shows $\rho_{\rm t} \propto
  S^{-3/2}$. This is the relation for a Euclidean population.}
\end{figure*}

\subsection{Parkes sky rates}\label{skyrate}
With the increased number of FRBs we update the all-sky rate estimates
for Parkes. The all-sky lower limit on the rate is
$4.7^{+2.1}_{-1.7}\times 10^3$~FRBs/($4\pi$~sr)/day. This is based on
the observed rate of $19$ events in $306$ days of observing with BPSR,
assuming the events occur within the full-width-half-power
field-of-view of the receiver, and extrapolating this to the entire
sky. The quoted uncertainties are $95\%$ Poisson uncertainties
\citep{Gehrels}. Additionally we update the fluence complete rate,
which is a more useful quantity when scaling FRB rates to other
telescopes and/or frequencies. Figure~\ref{figure:FRBfluence} shows
the observed peak flux density and observed widths of the FRB
population, with Parkes sensitivity and completeness regions
highlighted. Following \citet{Keanefluence} and considering those FRBs
in the fluence complete region we estimate a rate above $\sim
2\;\mathrm{Jy}\,\mathrm{ms}$ of $1.7^{+1.5}_{-0.9}\times
10^3$~FRBs/($4\pi$~sr)/day.

\subsection{Variable and transient source densities in the field of FRBs}
We essentially performed a targeted survey to search for significantly variable
and transient radio sources in the three of our FRB fields. We covered $\sim$0.15
deg$^{2}$ of sky for all fields with VLA at a sensitivity of
$\sim$100 $\upmu$Jy and $\sim$0.3 deg$^{2}$ of sky for all fields
with ATCA at a sensitivity of $\sim$300 $\upmu$Jy from 4~GHz to
8~GHz. We detected two sources in the VLA images of the field of FRB
151206 and one source in the ATCA images of the field of FRB 160102 to
vary significantly.

However, no radio transients were detected. The
significant variable source surface density for our survey is
$\rho_{\rm v}= 10^{+9.7}_{-5.4}$ deg$^{-2}$ (1-sigma Poisson
error). The Poisson uncertainties are calculated following
\citet{Gehrels}. The upper limits on the transient source density for
zero detections at 95\% confidence is given by $\rho_{\rm t}$ < 0.56
deg$^{-2}$ above the flux limit of 100 $\upmu$Jy. (Fig
\ref{figure:transient_density}).

\citet{bell} performed a search for variable sources in $\sim$0.3
deg$^{2}$ with comparable flux limits and at similar frequencies as
our search. They reported $\rho_{\rm v} = 3.3^{+7.5}_{-2.7}$
deg$^{-2}$ (1-sigma Poisson error) for significant variable
sources. We also compared our $\rho_{\rm v}$ with \citet{becker} and
\citet{mooley}. The results are presented in Fig
\ref{figure:source_density}. The flux density limit ($S_{\rm min}$) and
$\rho_{\rm v}$ for \citet{mooley} were scaled to 5.5 GHz from 3 GHz
using the relation $S_{ \min} \propto \nu^{\alpha}$ and $\rho_{\rm v}
\propto S_{\min}^{-1.5}$ where $\nu$ is the frequency and $\alpha$ is the
spectral index (which is assumed to be $- 0.7$). We find that the
surface density of significant variable sources is consistent within
the uncertainty estimates with surveys done in the past in non-FRB
fields. Consequently, we find no strong evidence that the FRBs
reported here are associated with the highly variable sources in the
fields, subject to the caveats that somewhat different variability
search criteria, different frequencies and different sensitivity
limits were used in the comparison surveys.

The probability of detecting $N$ variable sources in an area $A$ is
given by:
\begin{equation}
P(N) = \int_{0}^{\infty} P(N \mid \sigma) P(\sigma) d\sigma
\end{equation}
where, $\sigma$ is the variable source density, $P(\sigma)$ is the
prior probability for that variable, normalised such that $
\int_{0}^{\infty} P(\sigma) d\sigma =1$. We calculate the prior
probability using \citet{bell} as our control survey, which is given
by:
\begin{equation}
P(\sigma) = C \sigma^{N_{0}}~e^{-\sigma A_{0}}
\end{equation}
where C is the normalisation constant, $N_{0}$ and $A_{0}$ are the number
of highly variable source and the area covered in the control survey. We
use results from our VLA observations of FRB fields to compare with
the control survey because of their comparable sensitivities and found
that the probability of detecting two highly variable sources in a 
$\sim$0.15 deg$^{2}$ area of sky is 14.8$\%$. Currently with the
available data, we lack sufficient information to conclusively
associate any of these variable sources with FRB 151206 or FRB
160102. However, the detection of a known variable quasar in the field
of FRB 160102, the presence of variable AGN in the field of FRB 150418
\citep{Simon}, FRB 131104 \citep{radiocurve} and the persistent
variable radio source in the field of FRB 121102
\citep{VLAlocalisation} hint that FRBs might be related to AGN
activity in the host galaxy, however in the absence of a large FRB
population and their localisation, this remains speculative.

\section{Summary and conclusions}\label{conclusion}
We report the discoveries of four new FRBs in the SUPERB survey being
conducted with the Parkes radio telescope: FRB 150610, 151206, 
151230 and 160201. We have performed multi-messenger follow-up of these
using 2, 11, 12 and 8 telescopes respectively. No repeating radio pulses were detected in 103.1 hrs of radio follow-up. We continue to follow 
all SUPERB and bright HTRU FRBs in our ongoing SUPERB
observations.

\begin{table}
\caption{Comparison of the properties of the FRBs detected in SUPERB and the repeating FRB 121102. SUPERB FRBs are unresolved in time and show scattering unlike the repeater.}
\label{differences}
\centering
\begin{tabular}{|c|c|c|c|c|c|}
\hline 
Property & FRB 121102 & SUPERB FRBs \\
\hline
\hline
$\sim$100~MHz spectral features  & Yes &  1 of 5 sources\\
time resolved & Yes & No \\
range of spectral index  & $-$15 to $+$10 & $\sim$0 \\
scattering &  No & Yes \\
width & $3-9$~ms & $< 0.8-4$~ms \\
\hline
\end{tabular} 
\end{table}
A comparison of the repeating FRB with the published non-repeating
FRBs has been performed by \citet{Div2017}, who present
evidence that there are two distinct populations of FRBs -- repeating
and non-repeating -- based on the distribution of pulse fluences and
the amount of followup time for each source. The FRBs reported here
differ from FRB121102 (the repeating FRB) in a number of ways, as
shown in Table \ref{differences}. The pulses from the repeater are time resolved and
their pulse widths vary from 3$-$9 ms whereas the SUPERB FRBs are
unresolved (in time): the width is instead dominated by the effects of
DM smearing and scattering. This appears to provide further support
for the two source population conclusion of \citet{Div2017}.

With our larger sample of FRBs detected at Parkes, we have revisited
the FRB event rate and derived an updated all sky FRB rate of  $1.7^{+1.5}_{-0.9}\times 10^3$~FRBs/($4\pi$~sr)/day
above a fluence of  $\sim2\;\mathrm{Jy}\,\mathrm{ms}$. 
We have also computed the volumetric rate of FRBs for the 19 FRB 
sample using the fluence complete rate as our basis. We get volumetric rates in the range 2000 to 7000 Gpc$^{-3}$ $\rm yr^{-1}$ out to a redshift of $z~\sim~1$. 
This is consistent with volumetric rates for a range of transients (e.g. low luminosity long GRBs, short GRBs, NS-NS mergers, and supernovae (CC, Type Ia, etc)) \citep{Kulkarni,Totani}.

Our follow-up campaign of the reported FRBs yielded no multi-wavelength or multi-messenger counterparts and we have placed upper 
limits on their detection. We have also concluded that variability in the optical/radio images alone does not provide a reliable association with the FRBs. 
We encourage wide-field and simultaneous multi-wavelength observations of FRBs. In future, the detection of FRBs with an interferometer 
would be able to provide a robust host galaxy association.


\section*{Acknowledgements}

The Parkes radio telescope and the Australia Telescope Compact Array
are part of the Australia Telescope National Facility which is funded
by the Commonwealth of Australia for operation as a National Facility
managed by CSIRO. Parts of this research were conducted by the
Australian Research Council Centre of Excellence for All-sky
Astrophysics (CAASTRO), through project number CE110001020.
The GMRT is run by the National Centre for Radio Astrophysics 
of the Tata Institute of Fundamental Research. VLA is run by the National Radio Astronomy Observatory (NRAO).
NRAO is a facility of the National Science Foundation operated under cooperative agreement by Associated Universities, Inc. 
This work was performed on the gSTAR national
facility at Swinburne University of Technology. gSTAR is funded by
Swinburne and the Australian Government's Education Investment
Fund. 
This work is also based on data collected at Subaru Telescope,
which is operated by the National Astronomical Observatory of
Japan. We thank the LSST Project for making their code available as
free software at http://dm.lsstcorp.org. Funding from the European
Research Council under the European Union's Seventh Framework
Programme (FP/2007-2013) / ERC Grant Agreement n. 617199 (EP). Access
to the Lovell Telescope is supported through an STFC consolidated
grant. The 100-m telescope in Effelsberg is operation by the Max-Planck-Institut f{\"u}r Radioastronomie with
funds of the Max-Planck Society. The Sardinia Radio Telescope (SRT) is funded by the Department of
University and Research (MIUR), the Italian Space Agency (ASI), and the
Autonomous Region of Sardinia (RAS) and is operated as National Facility
by the National Institute for Astrophysics (INAF). TB and RWW are grateful to the 
STFC for financial support (grant reference ST/P000541/1).
Research support to IA is provided by the Australian Astronomical Observatory.
The ANTARES authors acknowledge the financial support of the funding agencies:
Centre National de la Recherche Scientifique (CNRS), Commissariat \`a
l'\'ener\-gie atomique et aux \'energies alternatives (CEA),
Commission Europ\'eenne (FEDER fund and Marie Curie Program),
Institut Universitaire de France (IUF), IdEx program and UnivEarthS
Labex program at Sorbonne Paris Cit\'e (ANR-10-LABX-0023 and
ANR-11-IDEX-0005-02), Labex OCEVU (ANR-11-LABX-0060) and the
A*MIDEX project (ANR-11-IDEX-0001-02),
R\'egion \^Ile-de-France (DIM-ACAV), R\'egion
Alsace (contrat CPER), R\'egion Provence-Alpes-C\^ote d'Azur,
D\'e\-par\-tement du Var and Ville de La
Seyne-sur-Mer, France;
Bundesministerium f\"ur Bildung und Forschung
(BMBF), Germany; 
Istituto Nazionale di Fisica Nucleare (INFN), Italy;
Stichting voor Fundamenteel Onderzoek der Materie (FOM), Nederlandse
organisatie voor Wetenschappelijk Onderzoek (NWO), the Netherlands;
Council of the President of the Russian Federation for young
scientists and leading scientific schools supporting grants, Russia;
National Authority for Scientific Research (ANCS), Romania;
Mi\-nis\-te\-rio de Econom\'{\i}a y Competitividad (MINECO):
Plan Estatal de Investigaci\'{o}n (refs. FPA2015-65150-C3-1-P, -2-P and -3-P, (MINECO/FEDER)), 
Severo Ochoa Centre of Excellence and MultiDark Consolider (MINECO), and Prometeo and Grisol\'{i}a programs (Generalitat
Valenciana), Spain; 
Ministry of Higher Education, Scientific Research and Professional Training, Morocco.
We also acknowledge the technical support of Ifremer, AIM and Foselev Marine
for the sea operation and the CC-IN2P3 for the computing facilities.
This work made use of data supplied by the UK Swift Science Data
Centre at the University of Leicester. This research has made use of
data, software and/or web tools obtained from the High Energy
Astrophysics Science Archive Research Center (HEASARC), a service of
the Astrophysics Science Division at NASA/GSFC and of the Smithsonian
Astrophysical Observatory's High Energy Astrophysics Division. This
work is based in part on data collected at Subaru Telescope, which is
operated by the National Astronomical Observatory of Japan. This paper
makes use of software developed for the Large Synoptic Survey
Telescope. We thank the LSST Project for making their code available
as free software at http://dm.lsstcorp.org. RPE/MK gratefully acknowledges support from 
ERC Synergy Grant "BlackHoleCam" Grant Agreement Number 610058

SB would like to thank Tara Murphy, Martin Bell, Paul Hancock, Keith
Bannister, Chris Blake and Bing Zhang for useful discussions.
\bibliographystyle{mnras}
\bibliography{references}

\onecolumn
\section*{Appendices}
\setcounter{subsection}{0}
\setcounter{table}{0}
\renewcommand{\thetable}{A\arabic{table}}
\renewcommand{\thesubsection}{\Alph{subsection}}

\subsection{FRB follow-up summary}\label{tables}
Tables \ref{table:FRB1details} to \ref{table:FRB4details} below
summarise all of the follow-up observations that have been performed for the four FRBs presented in this paper.

\begin{table}
\centering
\begin{tabular}{|c|c|c|c|c|}
  \hline
  Telescope & UTC & T post$-$burst & Tobs(sec) & Sensitivity limit\\
  \hline
  \hline
  ANTARES & 2015-06-10 05:26:58 & T$_{\rm FRB}$ &  T$_{\rm FRB}$ - day; T$_{\rm FRB}$ + day & Ref. Table \ref{tab:res_ANT1}  \\
  Parkes & 2017-06-08 03:22:10 & 728 days, 21:55:12 & 7200 & 466 mJy at 1.4~GHz\\
  Parkes & 2017-06-08 05:26:02 & 728 days, 23:59:04 & 7200 & 466 mJy at 1.4~GHz\\
  Parkes & 2017-06-08 07:38:43& 729 days, 2:11:45 & 7200 & 466 mJy at 1.4~GHz\\
  Parkes & 2017-06-08 09:43:47& 729 days, 4:16:49 & 7200 & 466 mJy at 1.4~GHz\\
  Parkes & 2017-06-08 11:48:30& 729 days, 6:21:32 & 7200 & 466 mJy at 1.4~GHz\\
    \hline
\end{tabular}
\caption{Multi-wavelength follow-up of FRB 150610 at ANTARES and Parkes. The sensitivity limits are specified for 10-sigma events with a width of 1ms at Parkes.  }
\label{table:FRB1details}
\end{table}

\begin{table}
\centering
\begin{tabular}{|c|c|c|c|c|}
  \hline
  Telescope & UTC & T post$-$burst & Tobs(sec) & Sensitivity limit\\
  \hline
  \hline
  ANTARES & 2015-12-06 06:17:52 & T$_{\rm FRB}$ &  T$_{\rm FRB}$ - day; T$_{\rm FRB}$ + day & Ref. Table \ref{tab:res_ANT1}  \\
  Parkes & 2015-12-07 07:52:39 & 1 day, 1:37:43 & 120 & 466 mJy at 1.4~GHz\\
  Parkes & 2015-12-07 07:55:28 & 1 day, 1:40:32 & 45.4 & 466 mJy at 1.4~GHz\\
  Parkes & 2015-12-07 07:57:16 & 1 day, 1:42:20 & 830 & 466 mJy at 1.4~GHz\\
  Lovell & 2015-12-07 09:49:43 & 1 day, 3:34:47 & 2982 & 350 mJy at 1.5~GHz\\
  TNT & 2015-12-07 12:00:27 & 1 day, 5:45:31 & 1500 & $r^{'}$ = 22.0\\
  SRT & 2015-12-07 13:57:30 & 1 day, 7:42:34 & 12177 & 1.7 Jy at 1.5~GHz\\
  e-Merlin & 2015-12 07-14:00:00 & 1 day, 7:45:04 & 18000 & 5 GHz - 204 $\upmu$Jy \\
  Effelsberg & 2015-12 07-14:36:10 & 1 day, 8:21:14 & 10800 & 240 mJy at 1.4~GHz\\
  SRT & 2015-12-07 15:00:00 & 1 day, 8:45:04 & 10800 & 1.7 Jy at 1.5~GHz \\
  UTMOST & 2015-12-08 04:26:42 & 1 day, 22:11:46 & 13500 & 11 Jy at 843~MHz \\
  Parkes & 2015-12-08 05:24:47 & 1 day, 23:09:51 & 1800 & 466 mJy at 1.4~GHz\\
  Parkes & 2015-12-08 05:55:27 &1 day, 23:40:31 & 1800 & 466 mJy at 1.4~GHz\\
  Parkes & 2015-12-08 06:26:07 & 2 days, 0:11:11 & 1800 & 466 mJy at 1.4~GHz\\
  Parkes & 2015-12-08 06:56:47 & 2 days, 0:41:51 & 1800 & 466 mJy at 1.4~GHz\\
  Parkes & 2015-12-08 07:27:28 & 2 days, 1:12:32 & 1800 & 466 mJy at 1.4~GHz\\
  Parkes & 2015-12-08 07:58:06 & 2 days, 1:43:10 & 550 & 466 mJy at 1.4~GHz\\
  e-Merlin & 2015-12-08 09:30:00 & 2 days, 3:15:04 & 32400 & 204 $\upmu$Jy at 5~GHz \\
  Lovell & 2015-12-08 18:09:16 & 2 days, 11:54:20 & 2985 & 350 mJy at 1.5~GHz \\
  VLA & 2015-12-08 19:38:01 & 2 days, 13:23:11 & 4497  & 70 $\upmu$Jy at 5.9~GHz  \\
  ATCA & 2015-12-09 01:58:35 & 2 days, 19:43:39 & 10800 & 200 $\upmu$Jy at 5.5~GHz  \\
  & & & & 280 $\upmu$Jy at 7.5~GHz  \\
  GMRT & 2015-12 09-04:15:00 & 2 days, 22:00:05 & 16200 & 180 $\upmu$Jy at 1.4~GHz  \\
  Lovell & 2015-12 09-17:02:04  & 3 days, 10:47:08 & 2990 & 350 mJy at 1.5~GHz \\
  VLA & 2015-12 10-18:45:22 & 4 days, 12:30:26 & 4498  & 70 $\upmu$Jy at 5.9~GHz\\
  TNT & 2015-12 11-11:57:22  & 5 days, 5:42:26 & 2940 & $r^{'}$ = 22.0\\
  VLA & 2015-12 12-19:22:22 & 6 days, 13:07:26 & 4498 & Badly affected by RFI\\
  VLA & 2015-12 14-19:44:22 & 8 days, 13:29:26 & 4498 & 70 $\upmu$Jy 5.9~GHz  \\
  Lovell & 2015-12 16-17:40:27 & 10 days, 11:25:31 & 2970 & 350 mJy at 1.5~GHz\\
  VLA & 2015-12-24 17:52:47 & 18 days, 11:37:51 & 4498 & 70 $\upmu$Jy at 5.9~GHz\\
  VLA & 2016-01-10 16:51:57 & 35 days, 10:37:01 & 4498& 70 $\upmu$Jy at 5.9~GHz\\
  VLA &2016-01-15 17:45:42 &40 days, 11:30:46 &4498& 70 $\upmu$Jy at 5.9~GHz \\
  VLA & 2016-03-06 13:39:23 & 91 days, 7:24:27 &4328 & 70 $\upmu$Jy at 5.9~GHz\\
  SRT & 2016-05-06 05:04:13 & 151 days, 22:49:17 & 10480 & 1.7 Jy at 1.5~GHz\\
  \hline
\end{tabular}
\caption{Multi-wavelength follow-up of FRB 151206 at 11 telescopes. The sensitivity limits are specified for 10-sigma events with a width of 1ms at Parkes, SRT, Lovell, Effelsberg and UTMOST.}
\label{table:FRB2details}
\end{table}

\begin{table}
\centering
  \begin{tabular}{|c|c|c|c|c|}
    \hline
    Telescope & UTC & T post$-$burst & Tobs(sec)&Sensitivity limit\\
    \hline
    \hline
    ANTARES & 2015-12-30 17:03:26 & T$_{\rm FRB}$ &  T$_{\rm FRB}$ - day; T$_{\rm FRB}$ + day & Ref. Table \ref{tab:res_ANT1} \\
    Zadko & 2015-12-30 18:03:21  & 00:59:55 & 7457 & $r$ $<$ 19.8\\
    Parkes & 2015-12-30 18:03:30 & 01:00:04 & 3616.01 & 466 mJy at 1.4~GHz \\
    Parkes & 2015-12-30 19:32:28 & 02:29:02 & 3618.11 & 466 mJy at 1.4~GHz \\
    SWIFT & 2015-12-30 23:14:45 & 06:11:19 & 2056.5 & 1.918 $\times$ 10$^{13}$ erg$^{-1}$cm$^{2}$ s$^{-1}$  \\
    DECam & 2015-12-31 07:11:17 & 14:07:51 & 900 & $u<$ 21.5\\
    DECam &2015-12-31 07:28:42	 &14:25:16 & 375 & $g<$ 22.5\\
    DECam &2015-12-31 07:37:22	 & 14:33:56 & 200 &  $r<$ 23.8\\
    DECam &2015-12-31 07:43:06	 & 14:39:40 & 750 & $i<$ 24.1 \\
    ATCA & 2015-12-31 14:15:45 & 21:12:19 & 28800 & 288 $\upmu$Jy at ~5.5 GHz \\
    & & & & 348 $\upmu$Jy at 7.5~GHz  \\
    Lovell & 2016-01-01 00:44:43 & 1 day, 7:41:17 & 7200 & 350 mJy at 1.5~GHz\\
    DECam & 2016-01-01 07:44:44 & 1 day, 14:41:18 & 200 & $g<$ 22.6\\
    Parkes & 2016-01-01 13:42:56 & 1 day, 20:39:30 & 3619.95 & 466 mJy at 1.4~GHz \\
    Parkes & 2016-01-01 14:43:39 & 1 day, 21:40:13& 3617.06 & 466 mJy at 1.4~GHz \\
    Parkes & 2016-01-01 15:44:19 & 1 day, 22:40:53 & 3617.06 & 466 mJy at 1.4~GHz \\
    Parkes & 2016-01-01 16:45:09 & 1 day, 23:41:43 & 3617.06 & 466 mJy at 1.4~GHz \\
    Parkes & 2016-01-01 17:45:49 & 2 days, 0:42:23 & 3617.06 & 466 mJy at 1.4~GHz \\
    Parkes & 2016-01-01 18:46:28 & 2 days, 1:43:02 & 3618.11 & 466 mJy at 1.4~GHz \\
    Parkes & 2016-01-01 19:47:09 & 2 days, 2:43:43 & 3617.06 & 466 mJy at 1.4~GHz \\
    Parkes & 2016-01-02 14:20:00 & 2 days, 21:16:34 & 3616.01 & 466 mJy at 1.4~GHz \\
    Parkes & 2016-01-02 15:20:38 & 2 days, 22:17:12 & 3618.11 & 466 mJy at 1.4~GHz \\
    Parkes & 2016-01-02 16:21:18 & 2 days, 23:17:52 & 3618.11 & 466 mJy at 1.4~GHz \\
    Parkes & 2016-01-02 17:22:10 & 3 days, 0:18:44 & 3616.01 & 466 mJy at 1.4~GHz \\
    Parkes & 2016-01-02 18:22:47 & 3 days, 1:19:21 & 3618.9 & 466 mJy at 1.4~GHz \\
    Parkes & 2016-01-02 19:23:28 & 3 days, 2:20:02 & 3618.11 & 466 mJy at 1.4~GHz \\
    Parkes & 2016-01-03 13:32:51 & 3 days, 20:29:25 & 3624.93 & 466 mJy at 1.4~GHz \\
    Parkes & 2016-01-03 14:33:38 & 3 days, 21:30:12 & 3618.11 & 466 mJy at 1.4~GHz \\
    Parkes & 2016-01-03 15:34:19 & 3 days, 22:30:53 & 3617.06 & 466 mJy at 1.4~GHz \\
    Parkes & 2016-01-03 16:35:09 & 3 days, 23:31:43 & 3617.06 & 466 mJy at 1.4~GHz \\
    Parkes & 2016-01-03 17:35:48 & 4 days, 0:32:22 & 3618.11 & 466 mJy at 1.4~GHz \\
    Lovell & 2016-01-03 22:41:31 & 4 days, 5:38:05 & 5580 & 350 mJy at 1.5~GHz\\
    Lovell & 2016-01-04 00:16:07 & 4 days, 7:12:41 & 1596 & 350 mJy at 1.5~GHz\\
    Parkes & 2016-01-04 14:54:30 & 4 days, 21:51:04 & 3616.01 & 466 mJy at 1.4~GHz \\
    Parkes & 2016-01-04 15:55:10 & 4 days, 22:51:44 & 3616.01 & 466 mJy at 1.4~GHz \\
    Parkes & 2016-01-04 16:56:00 & 4 days, 23:52:34 & 3616.01 & 466 mJy at 1.4~GHz \\
    Parkes & 2016-01-04 17:58:08 & 5 days, 0:54:42 & 3618.11 & 466 mJy at 1.4~GHz \\
    Parkes & 2016-01-04 18:58:49 & 5 days, 1:55:23 & 1258.03 & 466 mJy at 1.4~GHz \\
    Parkes & 2016-01-04 19:20:18 & 5 days, 2:16:52 & 3618.11 & 466 mJy at 1.4~GHz \\
    Parkes & 2016-01-05 14:40:00 & 5 days, 21:36:34 & 3616.01 & 466 mJy at 1.4~GHz \\
    Parkes & 2016-01-05 15:40:39 & 5 days, 22:37:13 & 3617.06 & 466 mJy at 1.4~GHz \\
    Parkes & 2016-01-05 16:41:42 & 5 days, 23:38:16 & 3623.88 & 466 mJy at 1.4~GHz \\
    Parkes & 2016-01-05 17:43:03 & 6 days, 0:39:37 & 3623.09 & 466 mJy at 1.4~GHz \\
    Parkes & 2016-01-05 18:52:06 & 6 days, 1:48:40 & 3619.95 & 466 mJy at 1.4~GHz \\
    Parkes & 2016-01-06 14:41:36 & 6 days, 21:38:10 & 3619.95 & 466 mJy at 1.4~GHz \\
    Parkes & 2016-01-06 15:42:16 & 6 days, 22:38:50 & 3619.95 & 466 mJy at 1.4~GHz \\
    Parkes & 2016-01-06 16:43:10 & 6 days, 23:39:44 & 3616.01 & 466 mJy at 1.4~GHz \\
    Parkes & 2016-01-06 17:43:47 & 7 days, 0:40:21 & 3618.9 & 466 mJy at 1.4~GHz \\
    Parkes & 2016-01-06 18:48:33 & 7 days, 1:45:07 & 3623.09 & 466 mJy at 1.4~GHz \\
    GMRT & 2016-01-06 18:30:00 &7 days, 1:26:34 & 15588 & 180 $\upmu$Jy at 1.4~GHz  \\
    Parkes & 2016-01-06 19:53:13 & 7 days, 2:49:47 & 1900.28 & 466 mJy at 1.4~GHz \\
    Subaru & 2016-01-07 11:23:19 & 7 days, 18:19:53 &  4200&  Refer table \ref{table2}\\
    Subaru & 2016-01-07 13:17:22 & 7 days, 20:13:56 &  3150& Refer table \ref{table2}\\\
    Subaru & 2016-01-07 15:12:39 & 7 days, 22:09:13 &  4200& Refer table \ref{table2}\\\
    Subaru & 2016-01-10 11:11:39  & 10 days, 18:08:13 & 3600& Refer table \ref{table2}\\\
    Subaru & 2016-01-10 13:03:29 & 10 days,  20:00:03 & 3600 & Refer table \ref{table2}\\\
    Subaru & 2016-01-10 15:07:20 & 10 days, 22:03:54 & 4080 & Refer table \ref{table2}\\\
    ATCA & 2016-01-11 11:36:55 & 11 days, 18:33:29 & 34440 & 288 $\upmu$Jy at 5.5~GHz \\
    & & & & 390 $\upmu$Jy at 7.5~GHz  \\
    \hline
  \end{tabular}
\end{table}
\begin{table}
\centering
  \begin{tabular}{|c|c|c|c|c|}
   \hline
    Telescope & UTC & T post$-$burst & Tobs(sec)&Sensitivity limit\\
    \hline
    \hline
    Subaru & 2016-01-13 11:21:54 &13 days, 18:18:28 &3600 & Refer table \ref{table2}\\\
    UTMOST & 2016-01-13 12:13:48 & 13 days, 19:10:22 & 27000 & 11 Jy at 843~MHz\\
    Subaru & 2016-01-13 13:12:52 &13 days, 20:09:26 & 3600& Refer table \ref{table2}\\\
    Subaru & 2016-01-13 15:13:35 &13 days, 22:10:09 & 3600& Refer table \ref{table2}\\\
    Lovell & 2016-01-14 00:03:12 & 14 days, 6:59:46 & 7200 & 350 mJy at 1.5~GHz \\
    Lovell & 2016-01-30 00:32:04 & 30 days, 7:28:38 & 7200 & 350 mJy at 1.5~GHz \\
    GMRT & 2016-02-17 19:30:00 & 49 days, 2:26:34 & 14400 & 180 $\upmu$Jy at 1.4~GHz   \\
    ATCA & 2016-02-24 09:48:15 & 55 days, 16:44:49 & 16500 & 240 $\upmu$Jy at 5.5~GHz   \\
    & & & & 252 $\upmu$Jy at 7.5~GHz  \\
    VLA & 2016-02-29 06:42:11 & 60 days, 13:38:45 & 4353 & 105 $\upmu$Jy at 5.9~GHz  \\
    GMRT & 2016-03-03 13:30:00 & 63 days, 20:26:34 & 14400 & 180 $\upmu$Jy at 1.4~GHz  \\
    VLA & 2016-03-04 06:26:25 & 64 days, 13:22:59 & 4353 & 84 $\upmu$Jy at 5.9~GHz  \\
    Lovell & 2016-03-18 18:34:51 & 79 days, 1:31:25 & 1965 & 350 mJy at 1.5~GHz \\
    SRT & 2016-05-10 17:58:43 & 132 days, 0:55:17 & 10350 & 1.7 Jy at 1.5~GHz\\
    \hline
  \end{tabular}
  \caption{Multi-wavelength follow-up of FRB 151230 at 12 telescopes. The sensitivity limits are specified for 10-sigma events with a width of 1ms at Parkes, SRT, Lovell and UTMOST.}
  \label{table:FRB3details}
\end{table}

\begin{table}
\centering
  \begin{tabular}{|c|c|c|c|c|}
    \hline
    Telescope & UTC & T post$-$burst & Tobs (sec) & Sensitivity limit\\
    \hline
    \hline
    ANTARES & 2016-01-02 08:28:38 & T$_{\rm FRB}$ &  T$_{\rm FRB}$ - day; T$_{\rm FRB}$ + day & Ref. Table \ref{tab:res_ANT1} \\
    Parkes & 2016-01-02 09:44:28 & 01:15:50 & 3618.11 & 466 mJy at 1.4~GHz  \\
    SWIFT & 2016-01-02 13:05:17 & 04:36:39 & 3582 & 1.434 $\times$ 10$^{13}$ erg$^{-1}$cm$^{2}$ s$^{-1}$  \\
    ATCA & 2016-01-03 02:42:45 & 18:14:07 & 14400 & 420 $\upmu$Jy  at 5.5~GHz \\
    & & & & 450 $\upmu$Jy at 7.5~GHz  \\
    Parkes & 2016-01-03 03:23:01 & 18:54:23 & 3624.93 & 466 mJy at 1.4~GHz\\
    Parkes & 2016-01-03 04:23:47 & 19:55:09 & 3618.9 &466 mJy at 1.4~GHz\\
    Parkes & 2016-01-03 05:50:16 & 21:21:38 & 3619.95 &466 mJy at 1.4~GHz\\
    Parkes & 2016-01-03 06:51:11 & 22:22:33 & 3624.93 &466 mJy at 1.4~GHz\\
    Parkes & 2016-01-03 08:15:34 & 23:46:56 & 3622.04 &466 mJy at 1.4~GHz\\
    Parkes & 2016-01-03 09:16:18 & 1 day, 0:47:39 & 3618.11 & 466 mJy at 1.4~GHz\ \\
    Parkes & 2016-01-03 10:16:59 & 1 day, 1:48:21 & 1556.87 & 466 mJy at 1.4~GHz\\\
    Parkes & 2016-01-04 10:18:14 & 2 days, 1:49:35 & 1179.39 & 466 mJy at 1.4~GHz\\\
    SWIFT & 2016-01-05 06:04:58 & 2 days, 21:36:20 & 1827 & 1.966 $\times$ 10$^{13}$ erg$^{-1}$cm$^{2}$ s$^{-1}$  \\
    Parkes & 2016-01-06 09:11:36 & 4 days, 0:42:57 & 3619.95 & 466 mJy at 1.4~GHz\\\
    Parkes & 2016-01-06 10:12:17 & 4 days, 1:43:38 & 896.27 & 466 mJy at 1.4~GHz\\\
    ATCA & 2016-01-11 05:34:35&8 days, 21:05:57 & 21060 & 330 $\upmu$Jy at 5.5~GHz \\
    & & & & 360 $\upmu$Jy at 7.5~GHz  \\
    UTMOST & 2016-01-13 06:43:00 &10 days, 22:14:22 & 16920 & 11 Jy at 843~MHz\\
    SWIFT & 2016-02-04 22:12:06 & 33 days, 13:43:28 & 3349 & 1.491 $\times$ 10$^{13}$ erg$^{-1}$cm$^{2}$ s$^{-1}$  \\
    GMRT & 2016-02-06 06:30:00 & 34 days, 22:01:22 & 14400 & 180 $\upmu$Jy  at 1.4~GHz \\
    ATCA & 2016-02-24 02:40:05 & 52 days, 18:11:27 & 23400 & 240 $\upmu$Jy  at 5.5~GHz \\
    & & & & 300 $\upmu$Jy at 7.5~GHz  \\
    VLA & 2016-02-26 17:50:17 & 55 days, 9:21:39 &4283  & 70 $\upmu$Jy at 5.9~GHz\\
    VLA & 2016-03-04 17:14:41 & 62 days, 8:46:03 & 4283 & 70 $\upmu$Jy at 5.9~GHz\\
    SRT & 2016-05-07 07:52:08 & 125 days, 23:23:30 & 7200 & 1.7 Jy at 1.5~GHz\\
    \hline
  \end{tabular}
  \caption{Multi-wavelength follow-up of FRB 160102 at 8 telescopes. The sensitivity limits are specified for 10-sigma events with a width of 1ms at Parkes, SRT and UTMOST.}
  \label{table:FRB4details}
\end{table}

\newpage

\subsection{Interferometric observational details and variability criteria}\label{observation_detail}
\begin{table}
\caption{Radio imaging observations performed with the ATCA, VLA and GMRT on the field of SUPERB FRBs. The table lists the number of epochs, area covered and primary \& secondary calibrators used for these observations. }
\label{table:imaging}
\resizebox{18cm}{!}{
\centering
\begin{tabular}{|c|c|c|c|c|c|c|c|c|c|c|c|c|}
\hline
 &  \multicolumn{4}{|c|}{ATCA} & \multicolumn{4}{|c|}{VLA} & \multicolumn{4}{|c|}{GMRT} \\
\hline
 & No. of epochs & Area (deg$^{2}$) & \multicolumn{2}{|c|}{PC \& SC}  & No. of epochs & Area (deg$^{2}$)& \multicolumn{2}{|c|}{PC \& SC}  & No. of epochs & Area (deg$^{2}$)&\multicolumn{2}{|c|}{PC \& SC}  \\
\hline
FRB 151260 & 1 & 0.05 & 1934$-$638 &1937$-$101 & 8 &0.05& 3C286 & J2355+4950 & 1 &0.05& 3C286 & 2011$-$067 \\
FRB 151230 & 3 & 0.05& 1934$-$638 & 0941$-$080 & 2 &0.05& 3C138 & J0943$-$0819 & 3 & 0.05 &3C48 &0943$-$083  \\
FRB 160102 & 3 &  0.2&1934$-$638 & 2240$-$260 & 2 &0.05 & 3C48 & J2248$-$3235 & 1 & 0.2& 3C48 &3C286 \\
\hline
\end{tabular}}
\end{table}
Table \ref{table:imaging} summarises the observations performed by the ATCA, VLA and GMRT on the field of SUPERB FRBs. For all detected sources the following statistics were used to test for variability using a method very similar to \citet{bell}. Firstly, the chi-square $\chi^{2}$ probability that the source is not variable is estimated with:
\begin{equation}
\chi^{2} =  \sum_{i=1}^{n} \frac{(S_{i}-\overline{S}_{\rm wt})^{2}}{\sigma_{i^{2}}}  
\end{equation}
where $S_{i}$ is the flux value in an epoch $i$, $\sigma_{i}$ is the
inverse of individual error in the flux measurement and
$\overline{S}_{\rm wt}$ is the weighted mean flux. Using $\chi^{2}$
distribution tables for $n -1$ degrees of freedom, a source is
classified as variable if $P < 0.001$ where $P$ is the probability
that $\chi^{2}$ is produced by chance. Additionally, the
de-biased modulation index is calculated using:
\begin{equation}
m_{\rm d}= \frac{1}{\overline{S}}\sqrt{\frac{ \sum_{i=1}^{n} (S_{i}-\overline{S})^{2}-\sum_{i=1}^{n}\sigma_{i}^{2}}{n}}
\end{equation}
where $\overline{S}$ is the mean flux density. 
Lastly, the fractional variability is computed using:
\begin{equation}
\Delta S = \frac{S_{\rm max} - S_{\rm min}}{ \bar{S}},
\end{equation}
where $S_{\rm max}$ and $S_{\rm min}$ are the maximum and minimum flux
densities for a source over $n$ epochs. A source is regarded to be a
significant variable if the $\chi^{2}$ is greater than threshold
$\chi^{2}_{\rm thresh}$ and $\Delta$S > 50$\%$, similar to
\citet{bell}.
\subsubsection{The Australian Telescope Compact Array }
The follow-up of three of the FRB fields was performed with the ATCA , using compact array broadband backend (CABB) \citep{Wilson2011} with a bandwidth of 2~GHz each centred at 5.5 GHz and 7.5 GHz to search for radio afterglows or variable sources associated with FRBs. The observations were done in a 42 pointing mosaic mode encompassing the localisation error radius of 7.5$\arcmin$. The data were reduced following the standard steps of imaging in \textit{miriad} \citep{miriad}. \textit{Aegean} \citep{Aegean} was used as a source finding and flux estimation software along with miriad tasks \texttt{IMSAD} and \texttt{IMFIT}. The images were searched for sources down to the threshold of 6-sigma in all ATCA data and a variability analysis (described above) was performed to identify variable sources. 
\subsubsection{The Karl G. Jansky VLA }
The VLA observations were performed in the 4~GHz to 8~GHz band with a centre frequency of 5.9~GHz. A seven pointing mosaic was done to encompass Parkes localisation error radius of 7.5$\arcmin$. The data reduction was performed using CASA \citep{CASA}. All sources detected above 7-sigma were monitored between the epochs to search for variable sources. We note here that the flux density scale using wide-band VLA mosaics is unreliable due to poorly constrained primary beam shape over the wide frequency band, however the flux scale is stable between epochs such that although the absolute flux scale of the mosaic images is wrong, the variability analysis will be correct. 
\subsubsection{The Giant Metrewave Radio Telescope }
The GMRT \citep{GMRT} observed the FRB fields at the center frequency of 1.4 GHz and bandwidth of 120 MHz. The data reduction was performed using the data reduction software AIPS \citep{AIPS}. \textit{Aegean} was used as source finding algorithm and a search for sources was performed down to 6-sigma noise level. 
\subsubsection{e-Merlin radio telescope}
The follow-up of FRB 151206 was also conducted by the e-Merlin telescope \citep{e-Merlin} with a bandwidth of 512 MHz centred on 5072.3 MHz. The data reduction was done using software AIPS \citep{AIPS} and a search for sources was performed down to 6-sigma noise limit. 
\subsection{Observational details and magnitude limits for non radio follow-ups.} \label{other}
\subsubsection{Thai National Telescope}
Optical follow up imaging was conducted on the field of FRB 151206 with the 2.4m Thai National Telescope (TNT), using the ULTRASPEC camera, with field of view 8$^{'}\times8^{'}$ \citep{dhillon}. Four tilings were observed on the night of 2015 December 7. Each tile observation consisted of 6 $r^{\prime}$-band images with exposure times of 60 seconds. The same 4 tiles were repeated 4 days later, enabling a comparative analysis of sources. The effective overlapping area observed on both occasions was 15$^{'} \times 15^{'}$, centred on 19:21:27, $-$04:07:35 (J2000). The estimated 5-sigma detection limits for both epochs were $r^{\prime}$ = 22.0. The variable sources detected are presented in Table \ref{table}.
\begin{table}
\centering
\caption{Optical variable sources detected by the Thai National Telescope (TNT) in the field of FRB 151206.}
\label{table}
\begin{tabular}{|c|c|c|c|c|}
\hline 
RA & DEC & r$^{\prime}$  mag  & $\Delta$ mag \\
\hline
\hline
19:21:28.47 & $-$04:08:50.5  &    17.8  &     +0.5\\
19:21:50.00 & $-$04:13:38.2  &    17.9  &   +0.2\\
19:21:01.30  &$-$04:12:00.4  &    18.3  &   +0.2\\
19:21:07.99 & $-$04:11:38.7  &    15.2  &   $-$0.1\\
\hline
\end{tabular} 
\end{table}
\subsubsection{Subaru Telescope}
The Hyper Suprime-Cam (HSC) data are reduced using HSC pipeline version 4.0.5 \citep{Bosch2017}, which is developed
based on the LSST pipeline \citep{,Ivezi, Axelrod,Juri2015},
in the usual manner including bias subtraction, flat-fielding, astrometry,
flux calibration, mosaicing, warping, coadding, and image subtraction.
The astrometric and photometric calibration is made relative to the Pan-STARRS1 \citep{panstars} with a 4.0 (24 pixel) aperture diameter.

For transient finding, the HSC pipeline adopts the frequently used image
subtraction algorithm developed by \citet{Alard} and \citet{Alard2};
an image with narrower point spread functions (PSFs) are convolved with
spatially varying kernels to match the wider PSFs of the other image, and
the image subtraction is made for the PSF-matched images. In the analysis,
we set the images taken on Jan 13 as the reference images and are subtracted
from the science images taken on Jan 7 and 10. The 5-sigma limiting magnitude on the variability
are estimated by $1000-4000$ apertures with a diameter being twice as large as the FWHM size
of PSF. The apertures are randomly sampled from positions without any detection
in the science and reference images and are locally sky subtracted.

Since the detected sources include many fakes, transient candidates are
further selected using their measured properties and the spatially varying PSF
and elongation of the difference images. We select the transient candidates
detected at least twice with the following detection criteria; (1) the detection
significance is higher than 5-sigma, (2) the PSF size is
between 0.8 and 1.3 of PSF size of the difference image, (3) the elongation is
larger than 0.65 of elongation of the difference image, and
(4) the residual of the subtraction of the PSF kernel from the detected source is
less than 3-sigma. The limiting magnitudes for Subaru observations are listed in Table \ref{table2}.
\begin{table}
\centering
\caption{Limiting magnitudes for Subaru observations of FRB 151230 field.}
\label{table2}
\begin{tabular}{|c|c|c|c|c|}
\hline 
Band & 2016-01-07 - 2016-01-13 &  2016-01-10 - 2016-01-13\\
\hline
\hline
HSC-G & >26.1                   & >26.3 \\
HSC-R & >25.8                  &>25.8 \\
HSC-I  & >26.00                    &>26.1\\
\hline
\end{tabular} 
\end{table}

\subsubsection{DECam}
The Dark Energy Camera \citep[DECam;][]{Diehl2012, Flaugher2012} is a
wide-field optical imager mounted at the primary focus of the 4-m
Blanco telescope at CTIO. Table \,\ref{table_exposures} and \ref{table_depth} summarises the details of
these observations and the limiting magnitudes

\begin{table}
\centering
  \caption{\small Details of the follow up of FRB 151230 performed with DECam, including the date and time of the observation, the filter used, the individual exposure time, and the number of exposures $N_{\rm exp}$ taken with a regular dithering pattern.}
  \begin{tabular}{l c c c }
    \hline
    Obs. time (UTC)	&	Filter	&	Exp (s)	& 	$N_{\rm exp}$\\
    \hline
    \hline
    2015-12-31, 07:11:17.1	&	$i$	&	180	&	5	\\
    2015-12-31, 07:28:42.5 & $r$  & 75  &   5  \\
    2015-12-31, 07:37:22.2  &  $g$   &   40    &  5   \\
    2015-12-31, 07:43:06.8  &  $u$   &   150  &  5   \\
    2016-01-01, 07:44:44.4 & $g$   &  40  & 5 \\
    \hline
  \end{tabular}
  \label{table_exposures}
\end{table}

\begin{table}
\centering
\caption{Detection limits (AB magnitudes) for sources detected in the DECam images for the field of FRB 151230 with significance reported in the last column. Refer to Table \ref{table_exposures} for more details about the observations.}
\begin{tabular}{l c c c }
\multicolumn{4}{c}{Detection limits}\\
\hline
 filter	& Date	&	$<$ mag (AB) & $N\sigma$	\\
 \hline
 \hline
$u$ & Dec 31\textsuperscript{st} & 21.52 &  5 \\
$g$ & Dec 31\textsuperscript{st} & 23.37 & 5\\
       &                                           & 22.55 & 10\\
$g$ & Jan 1\textsuperscript{st} & 23.53  & 5\\
       &                                         & 22.68  & 10\\
$r$ & Dec 31\textsuperscript{st}  & 23.84 &  5\\
$i$ & Dec 31\textsuperscript{st}  & 24.17  &  5\\

\hline
\end{tabular}
\label{table_depth}
\end{table}

The radio sources in ATCA and GMRT images for the FRB 151230 field
were compared with DECam optical image to look for optical
counterparts. Optical sources present above 5-sigma of the background
noise are considered to be a detection in each $u$-$g$-$r$-$i$
filter. We found that 52$\%$ of the radio sources have an optical
counterpart in at least one filter, for a search radius of 3
arcsec. This result of radio-to-optical source association is
consistent with the work of \citet{Huynh}.

\subsubsection{The Zadko Telescope}
The Zadko Telescope \citep{Zadko2017} is a 1 m f/4 Cassegrain telescope
situated in the state of Western Australia. The Zadko telescope has a
moderate field of view of $23^{\arcmin} \times 23^{\arcmin}$, so the
complete shadowing of the Parkes multi-beam receiver required 5-tile
images. 

\subsubsection{The ANTARES neutrino detector}
Searches for up-going track events in the ANTARES data have been optimised to give a 3-sigma discovery potential for one neutrino event in a search time window of $\rm{\Delta T}$ = [T$_0-$6 hr; T$\rm{_0+6~hr}$] within the ROI. For the four FRBs, the expected background event rate in a ROI of 2$^\circ$ is of the order of ${R_\mu \sim 5\cdot 10^{-8}}~\rm{event\cdot s^{-1}}$. Thus, the Poisson probability of observing zero event, knowing the background event rate, is $\ge$ 99$\%$ for any of the four FRBs.
Hence, the null result is compatible with the background expectation.  

The non detection of neutrino counterparts allows to derive upper limits at 90$\%$ confidence level on the neutrino fluence of the four FRBs  based on the instantaneous acceptance of ANTARES at the FRB trigger time: ${F_{\nu,90\%} < \int_{E_{\rm min}}^{E_{\rm max}}dN/dE_\nu\cdot E_\nu\cdot dE_\nu}$. Two generic neutrino energy spectra were considered and defined by a power law function ${dN/dE_\nu \propto E_\nu^{-\Gamma}}$ with spectral indices $\Gamma=\rm{1~and~2}$. The limits are then computed using a dedicated Monte Carlo simulation that takes into account the response of the detector at the FRB trigger time. The energy range  [$E_{\rm{min}}$; $E_{\rm{max}}$] corresponds to the 5-95$\%$ range of the energy distribution of the events in the optimised dataset. The results on the neutrino fluence upper limits for the two considered neutrino spectra are given in Table \ref{tab:res_ANT1}.
\begin{table}
\centering
\caption{Upper limits on the neutrino fluence, ${F_{\nu,90\%}}$, estimated for the 4 FRBs according to the instantaneous ANTARES sensitivity. The limits are given in  the energy range ${[E_{\rm min}-E_{\rm max}]}$ where 90$\%$ of the neutrino signal is expected.}
\begin{tabular}{|c|cc||cc|}
\hline
\multirow{3}{*}{FRB}& \multicolumn{2}{c||}{${\frac{dN}{dE_\nu} \propto E_\nu^{-2}}$} & \multicolumn{2}{c|}{${\frac{dN}{dE_\nu} \propto E_\nu^{-1}}$} \\ 
\cline{2-5}
\cline{2-5}
& ${F_{\nu,90\%}}$ & [${E_{\rm min};~E_{\rm max}}$] & ${F_{\nu,90\%}}$  & [${E_{\rm min};~E_{\rm max}}$] \\
& \scriptsize{$\rm{erg\cdot cm^{-2}}~(\rm{GeV\cdot cm^{-2}})$} &  \scriptsize{log$_{10}$[GeV]} & \scriptsize{$\rm{erg\cdot cm^{-2}}~(\rm{GeV\cdot cm^{-2}})$} & \scriptsize{ log$_{10}$[GeV] } \\ 
\hline

150610 & $3.2\cdot10^{-2}$ (20) & [3.4; 6.8] & 2.54 (1600) & [5.8; 7.9] \\ 

151206 & $1.8\cdot10^{-2}$ (11) & [3.6; 6.9] & 0.41 (250) & [5.8; 8.0]  \\ 

151230 & $1.8\cdot10^{-2}$ (11) & [3.2; 6.8] & 0.76 (470) & [5.8; 8.0]  \\ 

160102 & $2.6\cdot10^{-2}$ (16) & [3.6; 7.0] & 0.47 (290) & [5.8; 8.0]  \\ 

\hline
\end{tabular}
\label{tab:res_ANT1}
\end{table}
Constraints on the isotropic energy released in neutrinos can be set depending on the distance of the considered FRB: ${E_{\nu,90\%}^{\rm iso} = 4\pi D^2 \cdot F_{\nu,90\%}}/(1+z)$, where D is the effective distance travelled by the neutrinos. For the ${E_\nu^{-2}}$ spectral model, three FRB distance scenarios have been tested: a galactic environment with D = 50~kpc ($z\sim0$), a nearby extragalactic distance with ${D = 100}$ Mpc ($z\sim0.02$) and a cosmological scenario with D = D($z$) depending on the cosmological parameters and the maximum $z$ inferred from DM as listed in Table \ref{spects}. The cosmological distance, D($z$), travelled by the neutrinos from each FRB was computed from the Eq. 4 of \cite{ANTARES16} and found to be D($z$) = 6.61, 6.75, 3.67, 10.17~Gpc respectively. For the four FRBs, the ANTARES constraints given by ${E_{\nu,90\%}^{\rm iso}}$ are at the level of ${E_{\nu,90\%}^{\rm iso}}\sim10^{45}$, $10^{52}$ and $10^{55}$ erg, respectively for the three distance scenarios.
In particular, if these four FRBs are associated with neutrino emission following a ${E_\nu^{-2}}$ spectrum and with ${E_{\nu,90\%}^{\rm iso}}\gtrsim10^{52}$ erg, ANTARES excludes their origin at distance within 100~Mpc.

\newpage
\vspace{0.2cm}
\centering
\noindent \textsc{\large The ANTARES Collaboration}
\\
\vspace{0.2cm}
\scriptsize
	A.~Albert$^{32}$, 
	M.~Andr\'e$^{33}$, 
	M.~Anghinolfi$^{34}$, 
	G.~Anton$^{35}$, 
	M.~Ardid$^{36}$, 
	J.-J.~Aubert$^{37}$, 
	T.~Avgitas$^{38}$, 
	B.~Baret$^{38}$,
	J.~Barrios-Mart\'{\i}$^{39}$, 
	S.~Basa$^{40}$
% ...

\scriptsize
\vspace{0.2cm}
% ...
$^{32}$GRPHE - Universit\'e de Haute Alsace - Institut universitaire de technologie de Colmar, 34 rue du Grillenbreit BP 50568 - 68008 Colmar, France \\

$^{33}$Technical University of Catalonia, Laboratory of Applied Bioacoustics, Rambla Exposici\'o, 08800 Vilanova i la Geltr\'u, Barcelona, Spain \\
	$^{34}$INFN - Sezione di Genova, Via Dodecaneso 33, 16146 Genova, Italy \\
	$^{35}$Friedrich-Alexander-Universit\"at Erlangen-N\"urnberg, Erlangen Centre for Astroparticle Physics, Erwin-Rommel-Str. 1, 91058 Erlangen, Germany \\
	$^{36}$Institut d'Investigaci\'o per a la Gesti\'o Integrada de les Zones Costaneres (IGIC) - Universitat Polit\`ecnica de Val\`encia. C/  Paranimf 1, 46730 Gandia, Spain \\
	$^{37}$Aix Marseille Univ, CNRS/IN2P3, CPPM, Marseille, France \\
	$^{38}$APC, Univ Paris Diderot, CNRS/IN2P3, CEA/Irfu, Obs de Paris, Sorbonne Paris Cit\'e, France \\
	$^{39}$IFIC - Instituto de F\'isica Corpuscular (CSIC - Universitat de Val\`encia) c/ Catedr\'atico Jos\'e Beltr\'an, 2 E-46980 Paterna, Valencia, Spain \\
	$^{40}$LAM - Laboratoire d'Astrophysique de Marseille, P\^ole de l'\'Etoile Site de Ch\^ateau-Gombert, rue Fr\'ed\'eric Joliot-Curie 38,  13388 Marseille Cedex 13, France \\
% ...

\label{lastpage}
\end{document}